\documentclass[useAMS,usenatbib]{mn2e}
\usepackage{graphicx}
\def\apj{ApJ}
\def\apjs{ApJS}
\def\apjl{ApJL}
\def\aap{A\&A}
\def\aaps{A\&AS}
\def\aj{AJ}
\def\mnras{MNRAS}

\def\pasa{PASA}

\def\pasp{PASP}
\def\nat{Nature}
\def\sci{Science}

\def\araa{Ann.Rev.Astron.Astrophys.}

\def\fesc{\ifmmode f_{\rm esc} \else $f_{\rm esc}$\fi}
\def\fesclyc{$f_{\mathrm{esc}}$(LyC)}

\def\msun{M$_\odot$}

\def\oiil{[O~{\sc ii}]$\lambda 3727$}
\def\oiiil{[O~{\sc iii}]$\lambda 5007$}

\title[Lyman continuum leakage from star-forming galaxies]{Detection of high 
Lyman continuum leakage from four low-redshift compact star-forming galaxies}

\author[Y. I. Izotov et al.]{
Y. I. Izotov$^{1}$,\thanks{E-mail: izotov@mao.kiev.ua}
D. Schaerer$^{2,3}$, 
T. X. Thuan$^{4}$, 
G. Worseck$^{5}$, 
N. G. Guseva$^{1}$, 
\newauthor 
I. Orlitov\'a$^{6}$ \& A. Verhamme$^{2}$
\\
$^{1}$Main Astronomical Observatory, Ukrainian National Academy of Sciences,
27 Zabolotnoho str., Kyiv 03680, Ukraine\\
$^{2}$Observatoire de Gen\`eve, Universit\'e de Gen\`eve, 
51 Ch. des Maillettes, 1290, Versoix, Switzerland\\
$^{3}$IRAP/CNRS, 14, Av. E. Belin, 31400 Toulouse, France\\
$^{4}$Astronomy Department, University of Virginia, P.O. Box 400325, 
Charlottesville, VA 22904-4325, USA\\
$^{5}$Max-Planck-Institut f\"ur Astronomie, K\"onigstuhl 17, 69117 Heidelberg, Germany\\
$^{6}$Astronomical Institute, Czech Academy of Sciences, Bo\v cn{\'\i} II 1401, 141 00, Prague, Czech Republic
}

\begin{document}

\date{Accepted XXX. Received YYY; in original form ZZZ}

\pagerange{\pageref{firstpage}--\pageref{lastpage}} \pubyear{2012}

\maketitle

\label{firstpage}

\begin{abstract}
Following our first detection reported in Izotov et al. (2016), we 
present the detection of Lyman continuum (LyC) radiation of
four other compact star-forming galaxies observed with the Cosmic Origins 
Spectrograph (COS) onboard the {\sl Hubble Space Telescope} ({\sl HST}). 
These galaxies,
at redshifts of $z$ $\sim$ 0.3, are characterized by high 
emission-line flux ratios [O~{\sc iii}]$\lambda$5007/[O~{\sc ii}]$\lambda$3727  
$\ga$ 5. The escape fractions of the LyC radiation $f_{\rm esc}$(LyC) 
in these galaxies are in the range of $\sim$ 6~\% -- 13~\%, the highest values 
found so far in low-redshift star-forming galaxies. 
Narrow double-peaked Ly$\alpha$ emission lines are detected in the spectra of all four galaxies, compatible with predictions for Lyman continuum leakers.
We find escape fractions of Ly$\alpha$, $f_{\rm esc}$(Ly$\alpha$) 
$\sim$ 20\% -- 40\%, among the highest known for Ly$\alpha$ emitters (LAEs). 
Surface brightness profiles produced from the COS acquisition images reveal 
bright star-forming regions in the center and exponential discs in the 
outskirts with disc scale lengths $\alpha$ in the range $\sim$ 0.6 -- 1.4 kpc. 
Our galaxies are characterized by low metallicity, $\sim 1/8-1/5$ solar, 
low stellar mass $\sim$ (0.2 - 4) $\times$ 10$^9$ M$_\odot$,
high star formation rates SFR $\sim$ 14 -- 36  M$_\odot$ yr$^{-1}$,
and high SFR densities $\Sigma$ $\sim$ 2 -- 35 M$_\odot$ yr$^{-1}$ kpc$^{-2}$. 
These properties are comparable to those of high-redshift star-forming galaxies.
Finally, our observations, combined with our first detection reported in 
Izotov et al. (2016), reveal that a selection for compact star-forming 
galaxies showing high [O~{\sc iii}]$\lambda$5007/[O~{\sc ii}]$\lambda$3727 
ratios appears to pick up very efficiently sources with escaping Lyman 
continuum radiation: all five of our selected galaxies are LyC leakers.
\end{abstract}




\section{Introduction}\label{intro}

One of the key questions in observational cosmology is the identification of
the sources responsible for ionization of the Universe after the cosmic
Dark Ages, when the baryonic matter was neutral. 
Quasars ionize their surroundings, but are generally thought to be too rare 
to have contributed significantly 
to cosmic reionization \citep*{Fo12}. However, a new population of 
high-redshift, faint AGN candidates has recently
been  discovered  \citep{Gia15}, which, if confirmed, could have played an 
important role \citep{Madau15}.
The more commonly accepted picture is that galaxies are the main contributors
to reionization, although the currently identified sources are insufficient to fully ionize
the Universe by redshift $z \sim 6$ \citep*{S01,C09,Iw09,R13}. Fainter, low-mass star-forming galaxies (SFGs) 
below the current detection limit of observations are thought to be responsible for the
bulk of the ionizing radiation, since the observed UV luminosity function is very steep at
high-$z$ and hence faint galaxies are very numerous \citep*{O09,WC09,M13,Y11,B15a}. 

For galaxies to reionize the Universe, the escape fraction of their ionizing radiation has to be
high enough, typically of the order of 10--20 \% \citep[e.g.][]{R13,Robertson15}.
For example, \citet{D15} concluded that dwarf galaxies should provide, to the limit
of their observations with UV absolute magnitude $M$$_{\rm UV}$ $\sim$ $-$16 mag,
more than 20\% of the flux necessary to maintain ionization at $z$ = 5.7.
\citet{O09} have estimated a minimum escape fraction of the LyC radiation 
$f_{\rm esc}$(LyC) $>0.2$ required to maintain the Universe ionized at $z$ = 7.
On the other hand, \citet{Fi12,Fi15} estimated $f_{\rm esc}$(LyC) $<$ 13~\% at 
$z$ =6 if the luminosity function of galaxies extends to $M_{\rm UV} \sim -13$.
Others have examined the required average escape fraction from radiative
transfer models.
For example, \citet{K16} have calculated that  a constant $f_{\rm esc}$(LyC) of 14--22~\% 
is sufficient to reionize the Universe at $z> 5.5$. At $z$ $<$ 3.5 they find 
that $f_{\rm esc}$(LyC) 
can have values from 0 to 5~\%. However, a steep rise in $f_{\rm esc}$(LyC), 
of at least a factor of $\sim$ 3, is required between $z$ = 3.5 and 5.5, 
according to \cite{K16}. Other studies
argue for a slower increase of the escape fraction with redshift \citep[e.g.][]{HM12}. They model an increase based on available measurements (cited below in 
the text). So far, there is no physical explanation for an evolving escape 
fraction.

Although predicting the Lyman continuum escape fraction from galaxies {\em ab initio} is very challenging,
numerous simulations have tried to address this question, sometimes with contrasting  results.
For example, radiative hydro-dynamical simulations of galaxies at the epoch of 
cosmic reionization predict that the least massive galaxies have the highest  
\fesclyc\ \citep[e.g][]{R10,Y11,W14}, 
while other studies find relatively low escape fractions,  
\fesclyc\ $\approx 5$ \%, with no strong dependence 
on galaxy mass and cosmic time \citep[cf.][]{Y14,Ma15}, or
a declining escape fraction with decreasing galaxy mass \citep*{Gnedin08}.
According to \citet*{Kimm14} and \citet*{T15} simulations, the escape fraction 
of ionizing photons from low-mass galaxies is probably also strongly variable 
with time, due to the very bursty star formation histories in their low 
mass halos.  Stochastic escape from higher mass galaxies may
also be expected \citep*{P15,Ma15}. 

A different approach has recently been taken by  \citet{Sh16}, who considered a scenario in which ionizing
photons escape from regions with a high surface density $\Sigma$ of the star 
formation rate (SFR) as suggested by
some observations \citep{H11,B14}. They found that the fraction of the 
escaping ionizing radiation increases with redshift, 
reaching values of 5~\% -- 20~\% at $z$ $>$ 6, with the brighter galaxies 
having higher escape fractions.
Clearly, no consensus has been reached on this issue, and empirical data is needed.

Overall, searches for Lyman continuum leakers, both at high and low redshifts, 
have so far been difficult and largely unsuccessful.
Over the past, deep imaging studies at $z \sim 3$ have produced several candidate  leaking galaxies
\citep[e.g.][]{S01,Iw09,N11,Mosta13}, whereas other teams have only obtained stringent
upper limits \citep[e.g.][]{V10,B11}. \cite{V10b} has argued quite convincingly
that most of the LyC leaking candidates identified from imaging are explained 
by UV flux contamination by low-$z$ interlopers along the lines of sight
of the high-$z$ galaxies. Recent {\sl Hubble Space Telescope} ({\sl HST}) imaging 
from several groups appear to confirm this explanation, finding e.g.\ only one robust
 LyC detection from a sample of 21 candidates \citep{Si15,Mosta15}.
Currently, the most reliable Lyman continuum leaker detected at high redshift 
($z=3.212$) is the object {\em Ion2} by \cite{Va15} from deep imaging. It is 
confirmed and analyzed in detail by \cite{B16}. {\em Ion2} shows a high 
relative escape fraction $f^{\rm rel}_{\rm esc}$(LyC) = $0.64^{+1.1}_{-0.1}$ and
shares many properties with the Lyman continuum sources presented in this paper.

From the non-detections at high redshift, there are many empirical determinations of the upper limit 
of \fesclyc\ in the literature.
For example, \citet{R16} measured upper limits of $f_{\rm esc}$(LyC) $<$ 9.6~\% 
from a sample of SFGs at $z$ $\sim$ 1 selected to have H$\alpha$ equivalent 
widths EW(H$\alpha$) $>$ 200\AA, which are thought to 
be close analogs to the sources of reionization.
\citet{Sa15} put a 5$\sigma$ upper limit on the average $f_{\rm esc}$(LyC)
of $<$ 24~\% for H$\alpha$-emitting galaxies at $z$ = 2.2.
\citet{Gr15} found a low limit of  $<2$ \% for SFGs at $z=3.3$. 
\citet{B15a,B15b} have shown that $f_{\rm esc}$(LyC) cannot be in excess of 13\%
in galaxies at $z$ = 4 -- 5.
For the galaxies with $z> 3.06$ \citet{Si15} derived strong 1$\sigma$ limits on 
the relative escape fraction $f^{\rm rel}_{\rm esc}$(LyC) between 7~\% and 9~\%, 
while \citet{V10}
obtained an upper limit $f_{\rm esc}$(LyC) in the range 5~\% -- 20~\%.

Since direct observations of high-redshift galaxies are difficult because of 
their faintness, contamination by lower-redshift interlopers, and the increase 
of intergalactic medium (IGM) opacity \citep[e.g., ][]{V10,V12,Inoue14,Gr15}, 
one possible approach to bypass these difficulties is to 
identify local proxies of this galaxy population.
However, starburst galaxies at low redshifts are generally opaque to
their ionizing radiation \citep{L95,D01,Gr09}. This radiation with
small escape fractions $f_{\rm esc}$(LyC) of $\sim$1 -- 4.5~\% is directly 
detected only in four low-redshift galaxies. Two of these galaxies
were observed with the {\sl HST}/COS \citep{B14,L16}, one galaxy
with the {\sl Far Ultraviolet Spectroscopic Explorer} ({\sl FUSE}) \citep{L13}
and one galaxy with both the {\sl HST}/COS and {\sl FUSE} \citep{L13,L16}.
We note that {\sl FUSE} suffered from systematic errors for faint objects 
with flux densities $<$ few times 10$^{-15}$ erg s$^{-1}$ cm$^{-2}$ \AA$^{-1}$
as it is seen in Fig. 6 by \citet{FR07} and in Fig. 3 by \citet{Sh10}.
Knowing that their LyC fluxes are affected by these systematics, 
\citet{L13} tried their best to reduce these data. However, the 
{\sl FUSE} systematics (gain sag and scattered light) were never fully 
understood or calibrated, so that these challenging measurements might be 
beyond {\sl FUSE}'s capabilities.

On the other hand, low-mass compact galaxies at low redshifts $z < 1$ 
with very active star formation may be promising candidates for escaping 
ionizing radiation \citep*{Ca09,I11,JO13,S15}. The general characteristics of 
these galaxies is the presence of strong emission lines in the optical
spectra of their H~{\sc ii} regions powered by numerous O-stars, which produce
plenty of ionizing radiation.

  \begin{table}
  \caption{Coordinates, redshifts and O$_{32}$ ratios of selected galaxies
\label{tab1}}
\begin{tabular}{lrrcc} \hline
Name&R.A.(2000.0)&Dec.(2000.0)&$z$&O$_{32}$$^{\rm a}$ \\ \hline
J1152$+$3400&11:52:04.88&$+$34:00:49.88&0.3419&5.4\\ 
J1333$+$6246&13:33:03.96&$+$62:46:03.78&0.3181&4.8\\ 
J1442$-$0209&14:42:31.39&$-$02:09:52.03&0.2937&6.7\\ 
J1503$+$3644&15:03:42.83&$+$36:44:50.75&0.3557&4.9\\ 
\hline
\end{tabular}

$^{\rm a}$O$_{32}$ = [O {\sc iii}]$\lambda$5007/[O {\sc ii}]$\lambda$3727.

  \end{table}

  \begin{table*}
  \caption{Apparent magnitudes
\label{tab2}}
\begin{tabular}{lrrrrrrrrrrrcr} \hline
Name&\multicolumn{5}{c}{SDSS$^{\rm a}$}
&&\multicolumn{2}{c}{\sl GALEX}&&\multicolumn{4}{c}{{\sl WISE}} \\ 
    &\multicolumn{1}{c}{$u$}&\multicolumn{1}{c}{$g$}&\multicolumn{1}{c}{$r$}&\multicolumn{1}{c}{$i$}&\multicolumn{1}{c}{$z$}&&$FUV$&$NUV$&&\multicolumn{1}{c}{$W1$}&\multicolumn{1}{c}{$W2$}&\multicolumn{1}{c}{$W3$}&\multicolumn{1}{c}{$W4$}
\\
    &$u$(ap)&$g$(ap)&$r$(ap)&$i$(ap)&$z$(ap)&&&&&&&& \\ \hline
J1152$+$3400&20.43&20.28&19.74&20.63&19.79&&20.66&20.38&&17.00&15.96&12.65&\multicolumn{1}{c}{...} \\
            &20.72&20.60&20.06&20.86&20.01\\
J1333$+$6246&20.91&20.76&20.10&20.97&20.27&&20.72&20.97&&18.14&\multicolumn{1}{c}{...}&13.28&\multicolumn{1}{c}{...} \\
            &21.24&21.10&20.48&21.33&20.59\\
J1442$-$0209&20.40&20.30&19.48&20.73&19.77&&20.16&20.30&&17.14&15.83&11.64&8.29\\
            &20.76&20.61&19.83&21.03&20.01\\
J1503$+$3644&21.00&20.79&20.37&21.11&20.45&&21.01&20.76&&16.58&15.30&12.31&8.82\\
            &21.37&21.12&20.70&21.46&20.94\\ \hline
\end{tabular}

$^{\rm a}$$u,g,r,i,z$ are total modelled SDSS magnitudes. $u$(ap), $g$(ap), 
$r$(ap), $i$(ap), $z$(ap) are magnitudes within
the 3\arcsec\ diameter spectroscopic aperture for J1333$+$6246 and 
J1442$-$0209 and within the 2\arcsec\ diameter spectroscopic aperture for J1152$+$3400 and J1503$+$3644.

  \end{table*}

\begin{figure*}
\hbox{
\includegraphics[angle=0,width=0.24\linewidth]{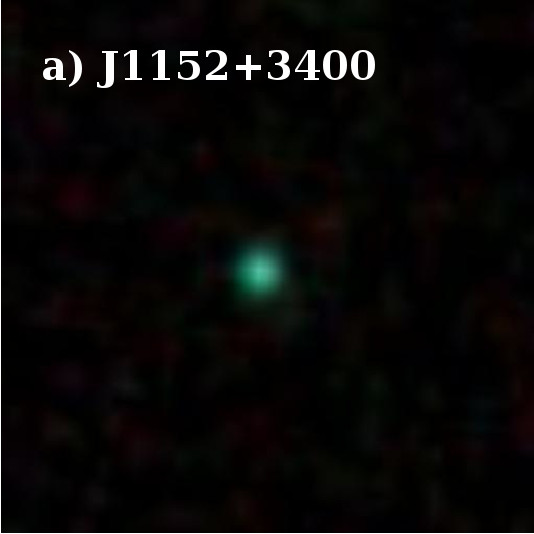}
\includegraphics[angle=0,width=0.24\linewidth]{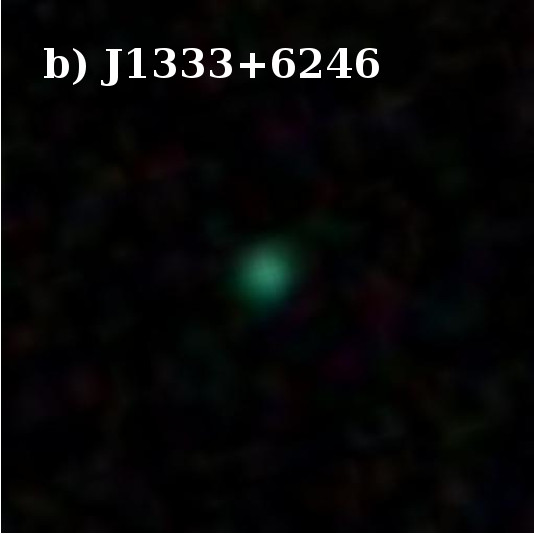}
\includegraphics[angle=0,width=0.24\linewidth]{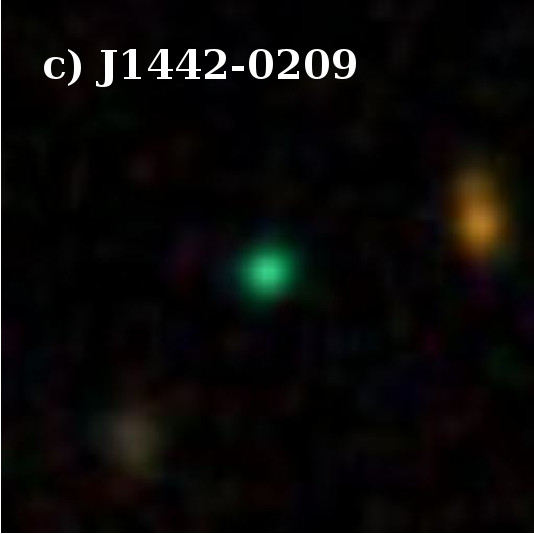}
\includegraphics[angle=0,width=0.24\linewidth]{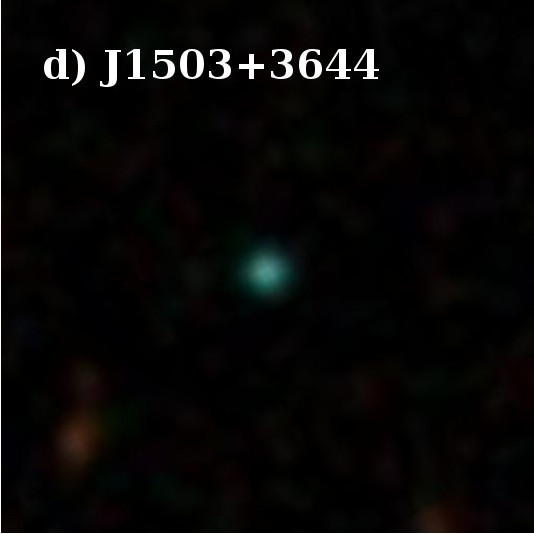}
}
\caption{25$''$$\times$25$''$ SDSS composite images of the LyC leaking galaxies.
\label{fig1}}
\end{figure*}

\citet{Ca09} used colour-colour diagrams to select a sample of ``Green Pea'' 
(GP) compact galaxies from the Sloan Digital Sky Survey (SDSS) named for their 
green colour. \citet{I11} instead used not only images but SDSS spectra as well
to produce a larger sample of Luminous Compact Galaxies (LCGs) 
with properties similar to GPs. They have shown that the colour of the compact
galaxy with active star formation is determined by the presence of strong
emission lines and therefore depends on the galaxy redshift. Therefore,
the GPs selected by \citet{Ca09} constitute a subsample of LCGs in a restricted
redshift range of $\sim$ 0.13 -- 0.3, while in general LCGs may have any colour.

In general LCGs and GPs are low-mass and low-metallicity galaxies 
\citep{I14a,I14b,I15}. Their stellar masses, SFRs and metallicities are similar
to those of high-redshift Lyman-alpha emitting (LAE) and Lyman-break 
galaxies (LBG) \citep{I15}. Many LCGs and GPs are characterised by 
high line ratios [O~{\sc iii}]$\lambda$5007/[O~{\sc ii}]$\lambda$3727 $\ga$ 5
(hereafter O$_{32}$), reaching values of up to 60 in some galaxies \citep{S15}.
Such high values may indicate that H~{\sc ii} regions are density-bounded 
allowing escape of ionizing radiation to the IGM, as suggested e.g.\ by 
\cite{JO13} and \cite{NO14}.

Based on these unique properties of LCGs and GPs, we have selected a sample of 
five galaxies for spectroscopic observations with the {\sl HST}, 
in conjunction with the Cosmic Origins Spectrograph (COS). Our aim is to detect
escaping ionizing radiation shortward of the Lyman continuum limit at
rest wavelengths $\la$ 912\AA\ and the Ly$\alpha$ emission line in these 
galaxies. Results for the first object, J0925+1403,
have been presented by \citet{I16} who found a high escape fraction 
$f_{\rm esc}$(LyC) = (7.8~$\pm$~1.1)~\% and detected a strong double-peaked 
Ly$\alpha$ emission line, indicating the high escape fraction 
$f_{\rm esc}$(Ly$\alpha$) of the radiation in this line.

In this paper we present the results of the {\sl HST}/COS observations for the
remaining four galaxies. The selection criteria are discussed in Section
\ref{sec:sample}. The {\sl HST} observations and data reduction are described
in Section \ref{sec:obs}. Extinction in the optical range and element abundances
are discussed in Section \ref{sec:ext}. Global characteristics of the galaxies
are derived in Section \ref{sec:global}. We derive surface brightness profiles 
in the near-ultraviolet (NUV) range in Section \ref{sec:sbp}. 
The reddening law in the UV range is
discussed in Section \ref{sec:red}. Ly$\alpha$ emission is considered in 
Section \ref{sec:lya}. The Lyman continuum detection and the corresponding
escape fractions are presented in Section \ref{sec:lyc}. 
Finally, we briefly discuss our results in Section \ref{sec:discuss} and 
summarize the main findings of the paper in Section \ref{summary}.

\section{The sample of compact star-forming galaxies and selection of
targets for {\sl HST} observations}\label{sec:sample}

   The spectroscopic data base of the SDSS Data Release 10 (DR10)
\citep{A14} was used to select a sample of compact SFGs applying the
following selection criteria \citep{I15}:
1) the angular galaxy radius on the SDSS images $R_{50}$ $\leq$ 3$''$, 
where $R_{50}$ is the galaxy's Petrosian radius within which 50\% of the 
galaxy's flux in the SDSS $r$ band is contained;
2) spiral galaxies were excluded; 
3) the emission-line ratio [O~{\sc iii}]$\lambda$4959/H$\beta$ is $\geq$ 1 to 
include only galaxies with high-excitation H~{\sc ii} regions; 
4) galaxies with AGN activity were excluded using line ratios. 
It was shown by \citet{I15} that these low-redshift compact SFGs obey 
similar mass-metallicity, luminosity-metallicity and mass-SFR relations 
as SFGs at higher redshifts ($z$ $\sim$ 2 -- 5).

Additional selection criteria were applied for the {\sl HST} observations 
\citep{I16}: 
1) a high equivalent width EW(H$\beta$) $>$ 100\AA\ of the H$\beta$ emission 
line in the SDSS spectrum; this ensures 
numerous hot O stars producing ionizing LyC radiation;
2) a sufficiently high brightness in the far-ultraviolet (FUV) and a high 
enough redshift ($z$ $\ga$ 0.3) to allow direct LyC observations with the 
COS; and
3) a high O$_{32}$ ratio $\ga 5$\footnote{In this paper, all line ratios are 
corrected for extinction derived from the Balmer decrement.}, 
which may indicate the presence of density-bounded H~{\sc ii} regions.

The five brightest galaxies in the FUV were selected for {\sl HST}/COS 
observations. Three of them were selected from the SDSS DR7 and two from the 
SDSS DR10. The data for the galaxy J0925$+$1403 have been presented in
\citet{I16}. The coordinates, redshifts, and O$_{32}$ ratios 
of the remaining four galaxies and 
their apparent magnitudes are shown in Tables \ref{tab1} and \ref{tab2}, 
respectively. Their SDSS images are presented in Fig. \ref{fig1}.
All galaxies exhibit a very compact structure. Since these LCGs are at 
redshifts of $\sim$ 0.3, their colours are green on SDSS composite images
and thus they can be classified as GP galaxies. 

\begin{figure*}
\begin{center}
\includegraphics[angle=-90,width=0.47\linewidth]{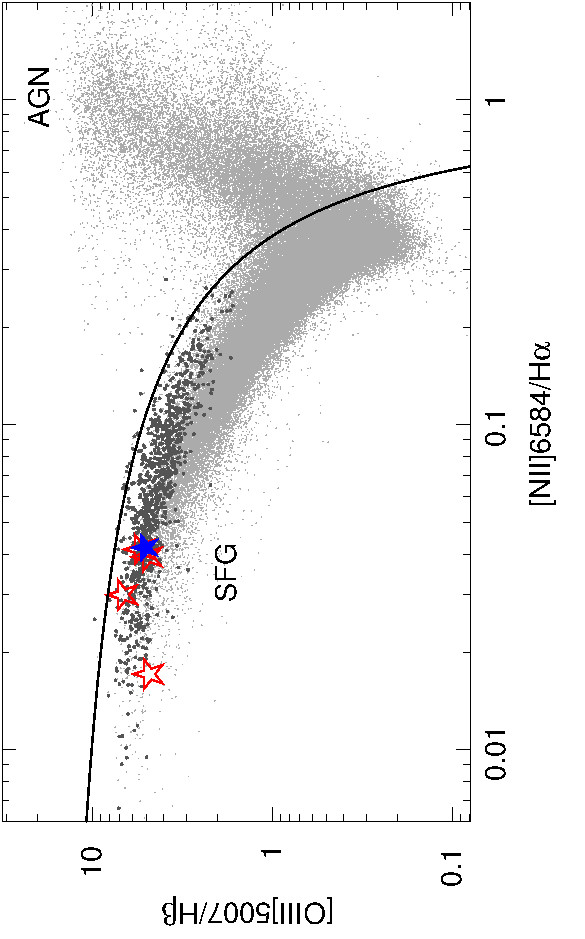}
\end{center}
\caption{The Baldwin-Phillips-Terlevich (BPT) 
diagram \citep{BPT81} for narrow emission-line galaxies.
The LyC leaking galaxies (this paper) and  J0925$+$1403 \citep{I16} are 
shown by red open stars and a blue filled star, respectively.
The Luminous Compact Galaxies \citep{I11} are represented by small dark-grey circles.
Also, plotted are the 100,000 emission-line galaxies from SDSS DR7 (cloud
of light-grey dots). The solid line by \citet{K03} separates 
star-forming galaxies (SFG) from active galactic nuclei (AGN).
\label{fig2}}
\end{figure*}

The location of the selected galaxies in the 
[O~{\sc iii}]$\lambda$5007/H$\beta$ -- [N~{\sc ii}]$\lambda$6584/H$\alpha$ 
diagnostic diagram \citep*{BPT81} is shown in Fig. \ref{fig2}. The solid line 
by \citet{K03} separates SFGs and active galactic nuclei (AGN). The selected 
galaxies, shown by stars, are located in the SFG region and thus their 
interstellar medium is ionized by hot stars in the star-forming regions. 

\begin{figure*}
\begin{center}
\includegraphics[angle=-90,width=0.47\linewidth]{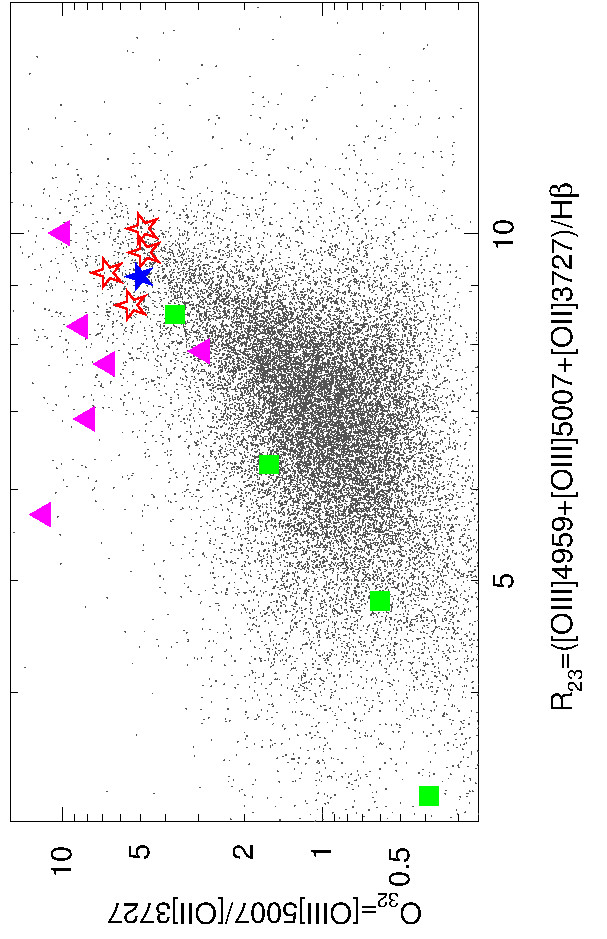}
\end{center}
\caption{The O$_{32}$ -- R$_{23}$ diagram for SFGs.
The quantity R$_{23}$ is the total flux of the strongest oxygen lines in the 
optical spectrum relative to H$\beta$. This quantity is used for easier 
comparison with high-redshift LAEs that are potentially leaking ionizing 
radiation \citep{NO14}, shown by magenta filled triangles. 
The location of LyC leaking galaxies corrected for extinction (this paper) 
and known low-redshift LyC leaking galaxies \citep{L13,B14,L16} are shown by red
open stars and green filled squares, respectively. The LyC leaking galaxy J0925$+$1403 
\citep{I16} is represented by the blue filled star. SDSS SFGs with 
extinction-corrected line intensities \citep{I14a} are shown by grey dots.
\label{fig3}}
\end{figure*}

We compare the locations in the O$_{32}$ -- R$_{23}$ 
(R$_{23}$=\{[O~{\sc ii}]3727+[O~{\sc iii}]4959+[O~{\sc iii}]5007\}/H$\beta$)
diagram of the selected galaxies (Fig. \ref{fig3}) with that of high-redshift
LAEs that are potentially leaking ionizing radiation
\citep{NO14}, and that of known low-redshift LyC leakers. It is seen that 
the selected galaxies shown by stars have properties similar to LAEs,
implying that they may be good LyC-leaking candidates. This has been
confirmed for J0925$+$1403 with $f_{\rm esc}$(LyC) of $\sim$8~\% \citep{I16}. 
On the other hand, the four low-$z$ LyC leakers \citep[Haro~11, Tol~1247$-$232,
J0921+4509 and Mrk~54, ][]{L13,B14,L16} with the lower $f_{\rm esc}$(LyC) of 
$\sim$1--4.5~\% are characterized by lower O$_{32}$ 
(green filled squares in Fig. \ref{fig3}).

\section{{\sl HST}/COS observations and data reduction}\label{sec:obs}

\begin{figure*}
\hbox{
\includegraphics[angle=-90,width=0.44\linewidth]{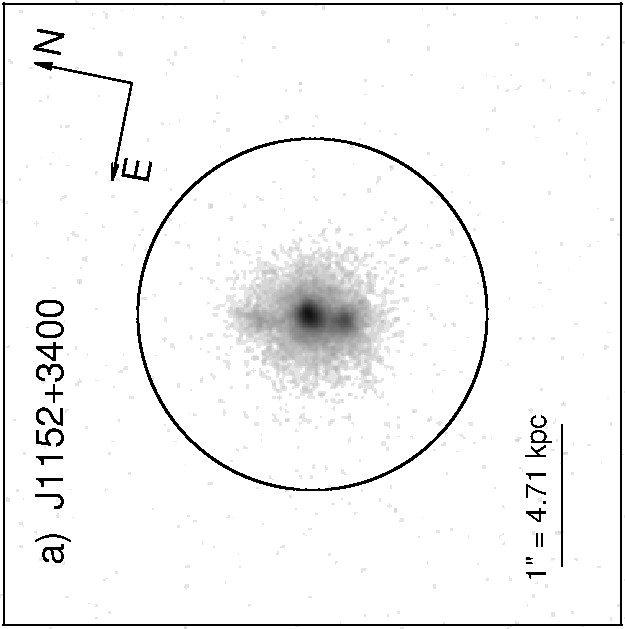}
\includegraphics[angle=-90,width=0.44\linewidth]{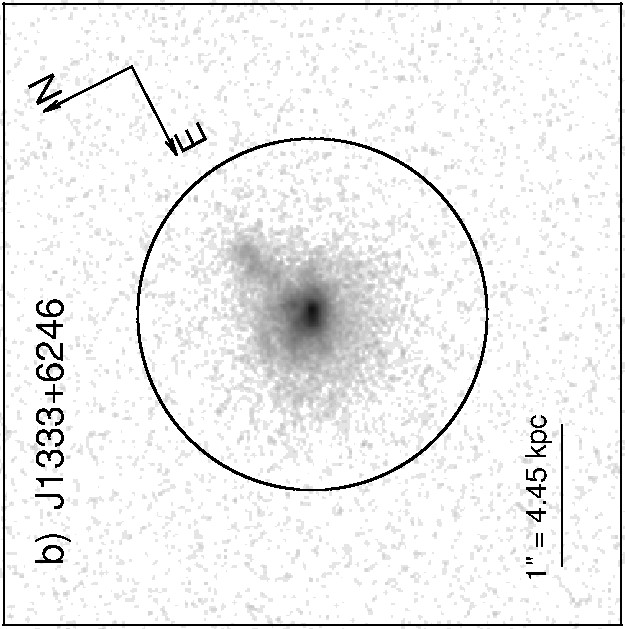}
}
\hbox{
\includegraphics[angle=-90,width=0.44\linewidth]{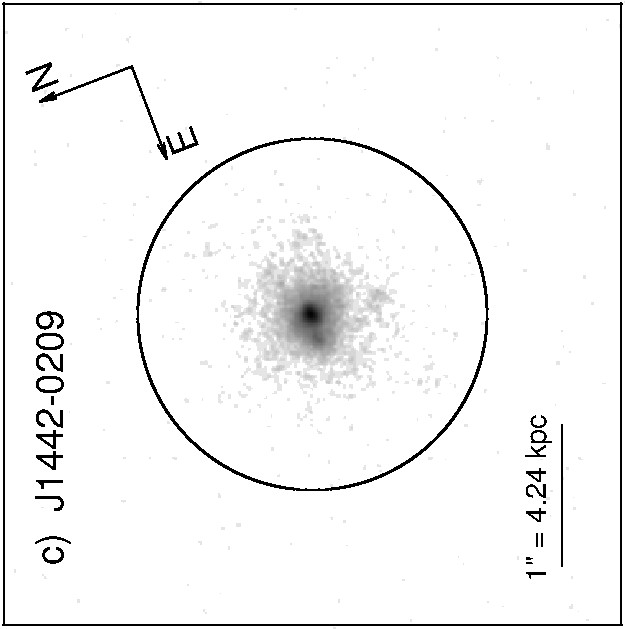}
\includegraphics[angle=-90,width=0.44\linewidth]{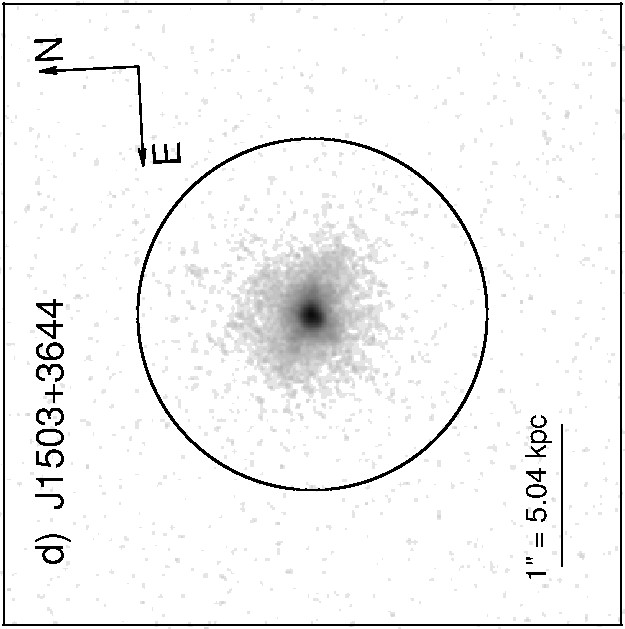}
}
\caption{The {\sl HST} COS/NUV
acquisition images of the LyC leaking galaxies in log surface brightness scale.
The COS spectroscopic aperture with a diameter of 2\farcs5 is shown by a circle.
The linear scale in each panel is derived adopting an angular size distance.
\label{fig4}}
\end{figure*}

The {\sl HST}/COS \citep{Gr12} observations were obtained in the course
of the program GO13744 (PI: T. X. Thuan).
The galaxies were first acquired using COS NUV images with the MIRRORA setting.
The galaxy region with the highest number of counts was automatically 
centered in the  2\farcs5 diameter spectroscopic aperture 
(Fig. \ref{fig4}). Although all galaxies show some structure with an extended 
low-surface-brightness (LSB) component, they are very compact and are localized
in the central part of the spectroscopic aperture, which is free of vignetting.

To obtain spectra we used two gratings, G140L and G160M, applying four 
focal-plane offset positions in each observation to minimize fixed-pattern 
noise and to patch grid-wire shadows and other detector blemishes 
(Table \ref{tab3}). 
The low-resolution G140L grating, with the central wavelength 1280\AA\ for
J1152$+$3400 \citep[same as for J0925$+$1403, ][]{I16} and 1105\AA\ for
the remaining galaxies, was used to obtain the spectrum, which 
includes the redshifted LyC emission. The medium-resolution G160M 
grating with two different central wavelengths was used to obtain spectra
with a spectral resolution sufficient to well resolve
the redshifted Ly$\alpha$ $\lambda$1216 \AA\ line and study its profile. 

  \begin{table*}
  \caption{{\sl HST}/COS observations \label{tab3}}
  \begin{tabular}{lcccc} \hline
\multicolumn{1}{c}{}&\multicolumn{1}{c}{}&\multicolumn{3}{c}{Exposure time (s)} \\ 
\multicolumn{1}{c}{Name}&\multicolumn{1}{c}{Date}&\multicolumn{3}{c}{(Central wavelength (\AA))} \\ 
    &    &MIRRORA&G140L&G160M \\ \hline
J1152$+$3400&2015-05-08& 558     &  2655&  1838,1875\\
            &          &         &(1280)&(1600,1623)\\
J1333$+$6246&2015-07-06& 982     &  2523&  1670,1667\\
            &          &         &(1105)&(1589,1623)\\
J1442$-$0209&2015-07-05& 491     &  2553&  1887,1890\\
            &          &         &(1105)&(1600,1623)\\
J1503$+$3644&2015-06-29& 808     &  5753&  1638,1617\\
            &          &         &(1105)&(1600,1623)\\ \hline
\end{tabular}
  \end{table*}

  \begin{table*}
  \caption{The observed LyC fluxes and photon counts \label{tab4}}
\begin{tabular}{lcccrc} \hline
 &Wavelength$^{\rm a}$&&\multicolumn{2}{c}{Counts}&Dark rate \\ 
Name& (\AA) &$I_{\rm obs}$(900\AA)$^{\rm b}$&Total$^{\rm c}$&Background&(s$^{-1}$)\\
\hline
J0925$+$1403&1120-1180&2.35$^{+0.21+0.00}_{-0.20-0.00}$&425&200.8$\pm$5.6&3.5$\times$10$^{-6}$\\
J1152$+$3400&1120-1180&4.28$^{+0.37+0.00}_{-0.34-0.00}$&287& 91.2$\pm$2.6&3.6$\times$10$^{-6}$\\
J1333$+$6246&1120-1180&0.83$^{+0.21+0.02}_{-0.20-0.02}$&124& 79.1$\pm$4.4&2.0$\times$10$^{-6}$\\
J1442$-$0209&1120-1178&1.98$^{+0.26+0.01}_{-0.27-0.01}$&192& 91.2$\pm$3.1&2.8$\times$10$^{-6}$\\
J1503$+$3644&1145-1180&1.60$^{+0.17+0.01}_{-0.16-0.01}$&277&119.8$\pm$4.6&2.0$\times$10$^{-6}$\\
\hline
  \end{tabular}

\hbox{$^{\rm a}$Wavelength range used for averaging.}

\hbox{$^{\rm b}$in 10$^{-17}$ erg s$^{-1}$cm$^{-2}$\AA$^{-1}$.}

\hbox{$^{\rm c}$Total count before background subtraction.}
 \end{table*}

The data were reduced with custom software specifically designed for faint 
{\sl HST}/COS targets \citep{W11,S12}. This gives more accurate spectra,
improving upon previous results on LyC leakage obtained with the default 
CALCOS pipeline \citep{B14}. 

The detector dark current in spectra of the four galaxies, which dominates 
the COS background, was subtracted  in the same way as for J0925$+$1403, as 
described in \citet{I16}.

One of us (G. Worseck) investigated the impact of background errors on 
the determination of the LyC flux. Monte Carlo simulations were run,
assuming that the background is a Gaussian variate
around the measured value and adopting an estimated uncertainty
equal to one standard deviation.
In Table \ref{tab4} we present the measured LyC fluxes with separate
statistical and systematic errors (first and second numbers for upper and 
lower 1$\sigma$ errors, respectively) for all four galaxies studied in this
paper and for J0925$+$1403 \citep{I16}. The total counts before background
subtraction and the background counts with 1$\sigma$ uncertainty are also
listed in the Table.
It is seen that the statistical error by far dominates the systematic error. 
Due to the large number of counts, the background error does not
have a large effect on the result, i.e. the measurements are not
background-limited. For J1333+6246 the background-subtracted signal is 
relatively small ($\sim$~45 counts), therefore the background uncertainty has 
a larger impact on the resulting LyC flux. The probabilities that the measured
counts are Poisson background fluctuations (i.e. the LyC flux from the
galaxy is zero) are 3$\times$10$^{-6}$ for J1333$+$6246 and 
$<$1$\times$10$^{-6}$ for the other galaxies. 

The diagnostic plots of the dark current monitoring 
programs\footnote{http://www.stsci.edu/$\sim$COS/fuv\_darks/dark\_vs\_time\_FUVA.png} 
show that the dark current is correlated with solar
activity. The declining radio flux during the year 2015 
(all targets were
observed between the end of March and the beginning of July 2015) indicates
decreasing solar activity. However, the dark rate still exhibits large
variations on short timescales, which could be either due to the variable 
Earth's magnetic field 
or due to short-time variations in the solar activity that are not
tracked well by the solar radio flux measurements. 

In the last column of Table \ref{tab4} we present the dark
rates in the LyC regions of our targets which
are roughly comparable to the dark rates in the monitoring programs,
but differ from them in three aspects:
1). Our dark rates include all pulse heights, whereas the diagnostic plots
from the monitoring cut very low and very high pulse heights.
2). Our dark rates are specifically for the LyC region. In particular, our
dark is estimated in the spectroscopic aperture, whereas the monitoring plots 
show the average over the detector which does neither account for gain sag in 
the aperture, nor local variations in the dark.
3). Our dark rates are estimated from post-processed dark-monitoring data
matching the conditions during the observations as closely as possible. 
A measurement of the true dark rate in the aperture requires post-processing
of dark monitoring data. Our estimated uncertainty in the dark varies with
the number of dark monitoring exposures matching the observing conditions
in a 3-month window around the observation date.

However, the information on the dark variation over timescales
of days is not available to us as the dark monitoring is only once per week 
(exposure 5$\times$1330 s, allocated during Earth occultation). The orbital 
telemetry data released to the public do not include several quantities that 
may be helpful for a deeper understanding of the observed dark variations 
(e.g. magnetic field strength and direction, thermospheric temperature and 
pressure). 
In summary, the impact of the
dark subtraction will not be large in this study. The time variation and
the relatively small intensity of the dark current will not appreciably
modify the LyC fluxes quoted here for our objects.

Airglow contamination (N~{\sc i} 1134\AA, N~{\sc i} 1200\AA, O~{\sc i} 1304\AA) 
was eliminated by considering only data taken in orbital night in the affected 
wavelength ranges. O~{\sc i}] 1356\AA\ line emission 
was negligible for all targets except 
for J1333$+$6246, for which it was excluded by considering only data taken at 
Earth limb angles $>$ 24\degr. For J1152$+$3400, J1333$+$6246 and 
J1442$-$0209, N~{\sc i} 1134\AA\ was considered negligible at all times, but to 
obtain a conservative flux estimate in the LyC region, we nevertheless 
substituted the wavelength range 1128 -- 1140\AA\ with shadow data. For 
J1503$+$3644, the only target with visible N~{\sc i} 1134\AA\ contamination, 
this was not possible due to detector blemishes falling into the wavelength 
range of interest during the Earth shadow passage, such that the wavelength 
range 1128 -- 1140\AA\ must be excluded from our analysis. For the targets 
observed in the 1105\AA\ setup we estimated and subtracted scattered 
geocoronal Ly$\alpha$ emission following Worseck et al. (2016, in press). The 
total amount of geocoronal scattered light varies with the specific day and 
nighttime fractions of the allocated {\sl HST} orbits, contributing a median 
scattered light flux in LyC range of (1.0 -- 2.5) $\times$ 10$^{-18}$ 
erg cm$^{-2}$ s$^{-1}$ \AA$^{-1}$. In practice, scattered light was only 
significant for J1333$+$6246, which has the lowest LyC flux and the smallest 
nighttime fraction during the observations (32~\%). For J1152$+$3400, for 
which geocoronal Ly$\alpha$ was not recorded in the spectrum, we verified 
that scattered light is negligible in the LyC range by comparing the total 
spectrum to the one restricted to the night time portion of the orbit.
Due to the small circular COS aperture open-shutter background cannot be 
estimated from the science exposures. 

While earthshine and zodiacal light 
are negligible, Galactic emission has been subtracted considering the 
exposure-time weighted average flux measured by \citet{M14} within a radius 
$r$ = 2\arcmin\ around our targets. The Galactic emission is small, but 
non-negligible (1~\% -- 8~\% of the measured LyC flux).

  \begin{table*}
  \caption{Extinction-corrected emission-line fluxes in SDSS spectra
\label{tab5}}
  \begin{tabular}{lrrrcrrcrrcrr} \hline
 & &\multicolumn{2}{c}{J1152+3400}
&&\multicolumn{2}{c}{J1333+6246}&&\multicolumn{2}{c}{J1442$-$0209}
&&\multicolumn{2}{c}{J1503+3644} \\ 
Line &\multicolumn{1}{c}{$\lambda$$^{\rm a}$} 
&\multicolumn{1}{c}{$I$$^{\rm b}$} &EW$_{\rm obs}$$^{\rm c}$ &
&\multicolumn{1}{c}{$I$$^{\rm b}$} &EW$_{\rm obs}$$^{\rm c}$ &
&\multicolumn{1}{c}{$I$$^{\rm b}$} &EW$_{\rm obs}$$^{\rm c}$ &
&\multicolumn{1}{c}{$I$$^{\rm b}$} &EW$_{\rm obs}$$^{\rm c}$  \\ \hline
Mg~{\sc ii}          &2796&  13.5$\pm$2.2&   8 &&  \multicolumn{1}{c}{...}    &  ...        &&  \multicolumn{1}{c}{...}    &  ... &&  33.3$\pm$3.8&  13       \\
Mg~{\sc ii}          &2803&   7.0$\pm$1.9&   4 &&  \multicolumn{1}{c}{...}    &  ...        &&  \multicolumn{1}{c}{...}    &  ... &&  19.9$\pm$3.0&   8       \\
$[$O~{\sc ii}$]$     &3727& 105.8$\pm$6.1& 134 &&130.2$\pm$11.&  77       && 93.6$\pm$5.1& 117&& 134.3$\pm$8.2& 170       \\
H12                  &3750&   5.0$\pm$3.4&   3 &&  \multicolumn{1}{c}{...}    &  ...        &&  6.2$\pm$2.1&   6&&   4.1$\pm$2.3&   4       \\
H11                  &3771&   6.0$\pm$2.8&   4 &&  \multicolumn{1}{c}{...}    &  ...        &&  5.7$\pm$2.1&   6&&   3.7$\pm$2.3&   3       \\
H10                  &3798&   7.5$\pm$2.5&   6 &&  \multicolumn{1}{c}{...}    &  ...        &&  6.2$\pm$2.0&   7&&   7.4$\pm$2.3&   9       \\
H9                   &3836&   8.7$\pm$2.3&   8 &&  \multicolumn{1}{c}{...}    &  ...        &&  6.9$\pm$2.1&   7&&   9.4$\pm$2.3&  12       \\
$[$Ne~{\sc iii}$]$   &3869&  44.5$\pm$3.2&  59 && 52.8$\pm$6.1&  42       && 52.2$\pm$3.2&  69&&  59.2$\pm$4.3&  85       \\
H8+He~{\sc i}        &3889&  20.1$\pm$2.5&  24 && 24.2$\pm$6.1&  17       && 20.0$\pm$2.3&  24&&  20.6$\pm$2.7&  28       \\
H7+$[$Ne~{\sc iii}$]$&3969&  28.0$\pm$2.7&  35 && 33.8$\pm$5.8&  32       && 34.3$\pm$2.7&  50&&  38.0$\pm$3.4&  56       \\
H$\delta$            &4101&  25.3$\pm$2.5&  32 && 27.3$\pm$5.7&  23       && 24.9$\pm$2.3&  40&&  29.6$\pm$2.9&  52       \\
H$\gamma$            &4340&  43.5$\pm$3.1&  68 && 47.1$\pm$6.3&  44       && 47.7$\pm$3.0&  94&&  51.1$\pm$3.8& 111       \\
$[$O~{\sc iii}$]$    &4363&   8.5$\pm$1.4&  14 && 17.1$\pm$3.7&  21       && 11.3$\pm$1.4&  23&&  12.4$\pm$1.7&  30       \\
He~{\sc i}           &4471&   4.0$\pm$1.1&   7 &&  \multicolumn{1}{c}{...}    &  ...        &&  \multicolumn{1}{c}{...}    &  ... &&   4.3$\pm$1.3&   9       \\
H$\beta$             &4861& 100.0$\pm$5.5& 198 &&100.0$\pm$8.5& 194       &&100.0$\pm$5.0& 312&& 100.0$\pm$6.0& 297       \\
$[$O~{\sc iii}$]$    &4959& 189.1$\pm$9.3& 407 &&208.7$\pm$15.& 434       &&205.7$\pm$9.2& 494&& 221.1$\pm$12.& 544       \\
$[$O~{\sc iii}$]$    &5007& 571.1$\pm$26.&1230 &&623.2$\pm$41.&1230       &&624.4$\pm$26.&1626&& 653.8$\pm$32.&1403       \\
He~{\sc i}           &5876&   8.9$\pm$1.1&  27 &&  9.0$\pm$2.5&  20       &&  5.5$\pm$1.0&  20&&  10.1$\pm$1.4&  35       \\
$[$O~{\sc i}$]$      &6300&   4.3$\pm$0.9&  17 &&  7.0$\pm$2.1&  21       &&  \multicolumn{1}{c}{...}    &  ... &&   4.4$\pm$0.9&  27       \\
$[$S~{\sc iii}$]$    &6312&   1.6$\pm$0.7&   6 &&  \multicolumn{1}{c}{...}    &  ...        &&  \multicolumn{1}{c}{...}    &  ... &&   1.8$\pm$0.7&  12       \\
H$\alpha$            &6563& 282.2$\pm$14.&1320 &&276.1$\pm$21.& 817       &&280.3$\pm$13.&1122&& 280.2$\pm$15.&1438       \\
$[$N~{\sc ii}$]$     &6583&  11.7$\pm$1.2&  53 &&  4.7$\pm$1.8&  21       &&  8.4$\pm$1.0&  38&&  11.1$\pm$1.4&  55       \\
$[$S~{\sc ii}$]$     &6717&  11.9$\pm$1.3&  56 &&  \multicolumn{1}{c}{...}    &  ...        &&  6.7$\pm$1.0&  22&&  10.5$\pm$1.3&  64       \\
$[$S~{\sc ii}$]$     &6731&  10.3$\pm$1.2&  47 &&  \multicolumn{1}{c}{...}    &  ...        &&  6.9$\pm$1.0&  21&&   8.9$\pm$1.2&  56       \\
He~{\sc i}           &7065&   5.0$\pm$0.8&  29 &&  \multicolumn{1}{c}{...}    &  ...        &&  \multicolumn{1}{c}{...}    &  ... &&   3.6$\pm$0.8&  20       \\
$[$Ar~{\sc iii}$]$   &7136&   6.4$\pm$0.9&  38 &&  \multicolumn{1}{c}{...}    &  ...        &&  \multicolumn{1}{c}{...}    &  ... &&   5.3$\pm$0.9&  50       \\ \\
$C$(H$\beta$)(int)  &&\multicolumn{2}{c}{0.085}&&\multicolumn{2}{c}{0.070}&&\multicolumn{2}{c}{0.085}&&\multicolumn{2}{c}{0.130}\\
$I_{\rm cor}$(H$\beta$)$^{\rm d}$       &&\multicolumn{2}{c}{30.8} &&\multicolumn{2}{c}{12.2} &&\multicolumn{2}{c}{38.9} &&\multicolumn{2}{c}{27.0} \\ \hline
  \end{tabular}

\hbox{$^{\rm a}$Rest-frame wavelength in \AA.}

\hbox{
$^{\rm b}$$I$ = 100$\times$$I_{\rm cor}$($\lambda$)/$I_{\rm cor}$(H$\beta$),
where $I_{\rm cor}$($\lambda$) and $I_{\rm cor}$(H$\beta$) are emission-line
fluxes, corrected for both the Milky Way and internal 
}
\hbox{
extinction.
}

\hbox{$^{\rm c}$Observed equivalent width in \AA.}

\hbox{$^{\rm d}$in 10$^{-16}$ erg s$^{-1}$ cm$^{-2}$.}

  \end{table*}

  \begin{table*}
  \caption{Physical conditions and chemical composition \label{tab6}}
  \begin{tabular}{lcccc} \hline
Galaxy &J1152+3400 &J1333+6246 &J1442$-$0209 &J1503+3644  \\ \hline
$T_{\rm e}$ ($[$O {\sc iii}$]$), K      & 13430$\pm$900       & 17780$\pm$2150       & 14580$\pm$830     & 14850$\pm$950        \\
$T_{\rm e}$ ($[$O {\sc ii}$]$), K       & 12980$\pm$820       & 15200$\pm$1720       & 13750$\pm$730     & 13910$\pm$830        \\
$T_{\rm e}$ ($[$S {\sc iii}$]$), K      & 12600$\pm$750       & 16320$\pm$1790       & 13480$\pm$690     & 13560$\pm$790        \\
$N_{\rm e}$ ($[$S {\sc ii}$]$), cm$^{-3}$&   310$\pm$300       &100$^{\rm a}$  &   710$\pm$620     &   280$\pm$360        \\ \\
O$^+$/H$^+$$\times$10$^{5}$             &1.60$\pm$0.32        &1.13$\pm$0.34        &1.21$\pm$0.19       &1.59$\pm$0.28          \\
O$^{2+}$/H$^+$$\times$10$^{5}$           &8.38$\pm$1.59        &4.59$\pm$1.34        &7.35$\pm$1.13      &7.40$\pm$1.27           \\
O/H$\times$10$^{5}$                     &9.98$\pm$1.62        &5.72$\pm$1.38        &8.56$\pm$1.15      &9.00$\pm$1.30          \\
12+log O/H                             &8.00$\pm$0.07        &7.76$\pm$0.11        &7.93$\pm$0.06      &7.95$\pm$0.06          \\ \\
N$^+$/H$^+$$\times$10$^{6}$             &1.16$\pm$0.17        &0.34$\pm$0.12        &0.74$\pm$0.10      &0.95$\pm$0.14          \\
ICF(N)$^{\rm b}$                &5.86                 &4.86                 &6.54               &5.38                   \\
N/H$\times$10$^{6}$                     &6.82$\pm$0.11        &1.64$\pm$0.60        &4.86$\pm$0.71      &5.11$\pm$0.79          \\
log N/O                                &~$-$1.17$\pm$0.10~~\, &~$-$1.54$\pm$0.19~~\, &~$-$1.25$\pm$0.09~~\,&~$-$1.25$\pm$0.09~~\,       \\ \\
Ne$^{2+}$/H$^+$$\times$10$^{5}$          &1.67$\pm$0.35        &0.89$\pm$0.27        &1.52$\pm$0.25       &1.63$\pm$0.30          \\
ICF(Ne)$^{\rm b}$               &1.07                 &1.09                 &1.06                &1.09                   \\
Ne/H$\times$10$^{5}$                    &1.79$\pm$0.42        &0.97$\pm$0.33        &1.60$\pm$0.29       &1.77$\pm$0.37          \\
log Ne/O                               &~$-$0.75$\pm$0.12~~\, &~$-$0.77$\pm$0.18~~\, &~$-$0.73$\pm$0.10~~\,&~$-$0.71$\pm$0.11~~\,     \\ \\
S$^{+}$/H$^+$$\times$10$^{6}$            &0.29$\pm$0.04        &  ...               &  ...              &0.22$\pm$0.03        \\
S$^{2+}$/H$^+$$\times$10$^{6}$           &1.45$\pm$0.68        &  ...               &  ...              &1.29$\pm$0.54        \\
ICF(S)$^{\rm b}$                &1.43                 &  ...               &  ...              &1.34                 \\
S/H$\times$10$^{6}$                     &2.48$\pm$0.97        &  ...               &  ...              &2.03$\pm$0.73        \\
log S/O                                &~$-$1.60$\pm$0.18~~\, &  ...               &  ...              &~$-$1.65$\pm$0.17~~\,    \\ \\
Ar$^{2+}$/H$^+$$\times$10$^{7}$          &3.57$\pm$0.63        &  ...               &  ...              &2.57$\pm$0.49          \\
ICF(Ar)$^{\rm b}$               &1.15                 &  ...               &  ...              &1.14                   \\
Ar/H$\times$10$^{7}$                    &4.12$\pm$0.72        &  ...               &  ...              &2.92$\pm$0.56          \\
log Ar/O                               &~$-$2.38$\pm$0.10~~\, &  ...               &  ...              &~$-$2.49$\pm$0.10~~\,    \\ \hline
\end{tabular}

\hbox{$^{\rm a}$Assumed value.}
\hbox{$^{\rm b}$Ionization correction factor.}
  \end{table*}

\section{Extinction in the optical range and element abundances}\label{sec:ext}

The observed decrement of several hydrogen Balmer emission lines 
in the SDSS spectra
is used to correct the line fluxes for reddening according
to \citet*{ITL94} and adopting the reddening law by \citet*{C89}.
The correction for the Milky Way and internal extinction was done 
separately as described in \citet{I16}.
The extinction coefficient $C$(H$\beta$) derived 
from the comparison of the observed and theoretical case B hydrogen Balmer 
decrements corresponds to the extinction  
$A$(H$\beta$) = 2.512~$\times$~$C$(H$\beta$) at the H$\beta$ wavelength.
To derive the extinction $A(V)$ in the $V$ band \citet{I16} approximated the 
ratio $C$(H$\beta$)/$A(V)$ as a function of $R(V)$=$A(V)$/$E(B-V)$.

The extinction-corrected emission-line fluxes relative to the H$\beta$ 
emission line fluxes and the observed equivalent widths are shown 
in Table \ref{tab5}. 
The Table also gives the internal extinction 
coefficients $C$(H$\beta$) and the
H$\beta$ emission-line fluxes $I_{\rm cor}$(H$\beta$)
corrected for both the Milky Way and internal extinctions. 

We use the emission line intensities (Table \ref{tab5}) and the direct 
$T_{\rm e}$-method to derive electron temperatures and electron number
densities, ionic and total element abundances in
the interstellar medium (ISM) of the galaxies as described in \citet{I06}.
The derived temperatures and element abundances 
are shown in Table \ref{tab6}. The oxygen abundances of all galaxies are
in the narrow range of $\sim$ 7.8 -- 8.0, i.e. $\sim$1/8 -- 1/5 that of the 
Sun, if the solar abundance of \citet{A09}, 12 + log O/H = 8.69
is adopted. These values are similar to the value of
$\sim$ 7.91 found for J0925$+$1403 \citep{I16}. The ratios of other
element abundances to oxygen abundance are in the range obtained for
dwarf emission-line galaxies \citep[e.g. ][]{I06}.

\begin{figure*}
\hbox{
\includegraphics[angle=-90,width=0.45\linewidth]{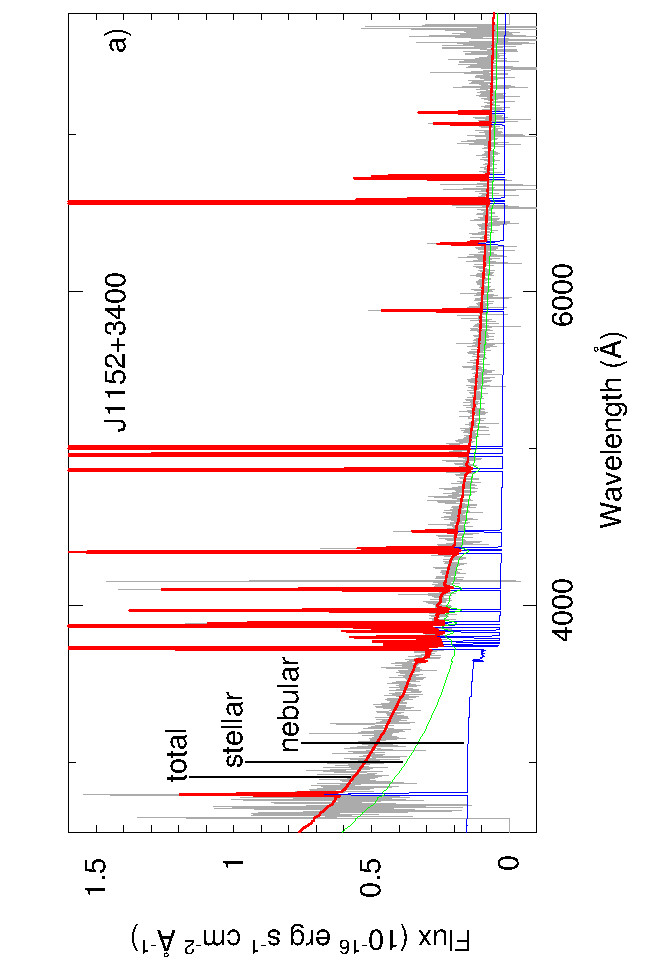}
\hspace{0.5cm}\includegraphics[angle=-90,width=0.45\linewidth]{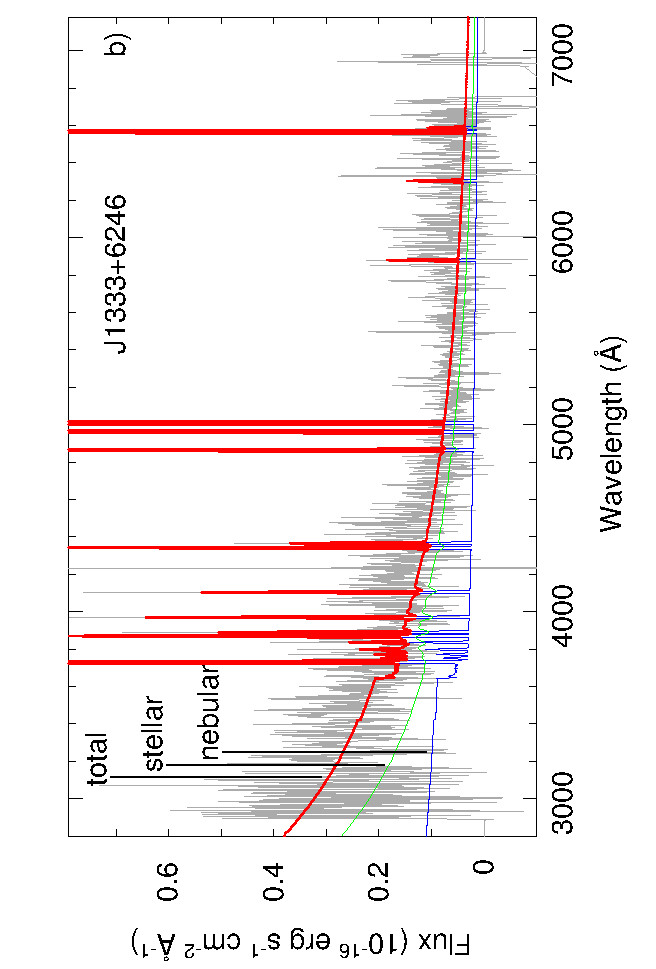}
}
\hbox{
\includegraphics[angle=-90,width=0.45\linewidth]{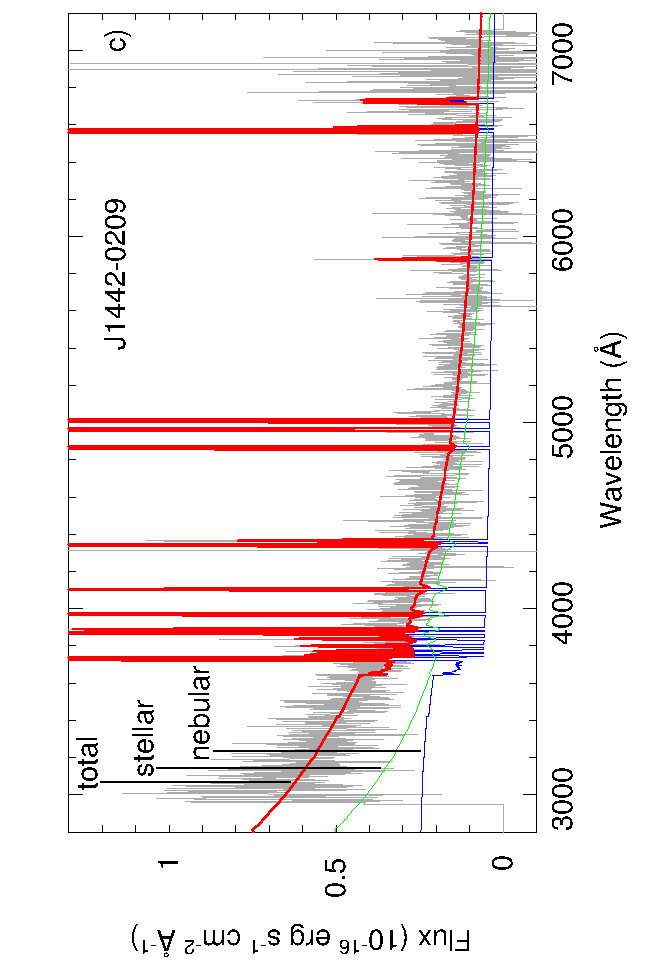}
\hspace{0.5cm}\includegraphics[angle=-90,width=0.45\linewidth]{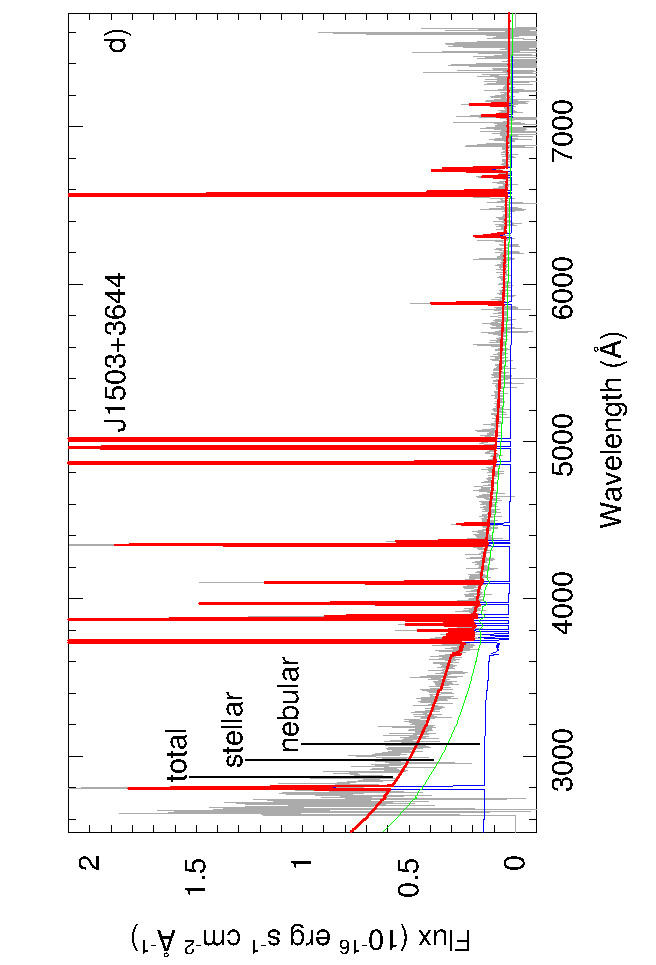}
}
\caption{SED fitting of the optical spectra of LyC leaking galaxies.
The rest-frame extinction-corrected spectra are shown by grey lines.
The stellar, nebular, and total modelled SEDs are shown by thin green, thin blue
and thick red lines, respectively, and are labelled.
\label{fig5}}
\end{figure*}

  \begin{table*}
  \caption{Global parameters \label{tab7}}
  \begin{tabular}{lccccccc} \hline
Name&12+logO/H&  $D$$^{\rm a}$   &log $M_\star$$^{\rm b}$ &SB age&SFR&$\alpha$$^{\rm c}$
&$\Sigma$$^{\rm d}$\\
    &         & (Mpc)  &(log M$_\odot$)& (Myr)&(M$_\odot$ yr$^{-1}$)&(kpc)&
(M$_\odot$ yr$^{-1}$kpc$^{-2}$) \\ \hline   
J1152$+$3400&8.00$\pm$0.07&1888&9.59&4.1&39&0.59& 35.5\\
J1333$+$6246&7.76$\pm$0.11&1736&8.50&1.6&14&1.44&~\,2.2\\
J1442$-$0209&7.93$\pm$0.06&1582&8.96&3.4&36&0.86& 15.5\\
J1503$+$3644&7.95$\pm$0.06&1975&8.22&4.0&38&1.34&~\,6.8\\
\hline
  \end{tabular}

\hbox{$^{\rm a}$Luminosity distance.}

\hbox{$^{\rm b}$$M_\star$ is the stellar mass.}

\hbox{$^{\rm c}$Exponential disc scale length.}

\hbox{$^{\rm d}$$\Sigma$ = 12.6 $M_\odot$ yr$^{-1}$kpc$^{-2}$ for J0925$+$1403.}
  \end{table*}

\section{Global parameters of the galaxies}\label{sec:global}

We use spectral energy distribution (SED) fits to derive galaxy stellar masses.
The fits were performed for the SDSS spectra over the 
entire observed spectral range of 3900--9200~\AA\ for J1333$+$6246 and 
J1442$-$0209 selected from the SDSS DR7 
\citep[same as for J0925$+$1403, ][]{I16}, and over 3600--10300~\AA\ for 
J1152$+$3400 and J1503$+$3644 selected from the SDSS DR10.

In the galaxies studied here, with rest-frame H$\beta$ 
equivalent widths EW(H$\beta$) $\sim$ 150 -- 200 \AA, nebular continuum 
and line emission are strong and their contributions are removed
when determining the stellar mass.

To fit the SEDs we carried out a series of Monte Carlo simulations
using the technique described e.g. in 
\citet{G07}, \citet{I15} and \citet{I16}.
The best solutions can be found for different combinations of evolutionary 
tracks, stellar atmosphere models and initial mass functions. 
As shown by \citet{I16}, all these solutions provide almost equally good fits at
rest-frame wavelengths greater 912\AA\ and show small variations of the LyC. 
Therefore, in this paper, for the sake of definiteness, we have adopted a 
Salpeter IMF \citep{S55}, Geneva evolutionary tracks \citep{M94} 
of non-rotating stars and a 
combination of stellar atmosphere models \citep{L97,S92}. The optical 
galaxy spectra with the overlaid stellar (thin green lines), nebular
(thin blue lines) and total stellar and nebular (thick red lines) SEDs are 
shown and labelled in Fig. \ref{fig5}.

The observed fluxes were transformed to
luminosities adopting luminosity distances derived with the cosmological 
calculator \citep[NED,][]{W06}, based on the cosmological 
parameters $H_0$=67.1 km s$^{-1}$Mpc$^{-1}$, $\Omega_\Lambda$=0.682, 
$\Omega_m$=0.318 \citep{P14}. Stellar masses and starburst ages derived from 
the SED fitting of the SDSS optical spectra are presented in 
Table \ref{tab7}. We also derived extinction-corrected 
absolute AB SDSS $g$-band and {\sl Galaxy Evolution Explorer} ({\sl GALEX}) 
FUV magnitudes not shown in the
Table. They are in the range $-$20.4 - $-$21.3 mag, while the magnitudes
non-corrected for extinction are $\sim$ 0.2 -- 0.3 mag and $\sim$ 1 mag 
fainter in the $g$ and FUV bands, respectively.
We note that the photometric data and UV spectra were not used in the SED 
fitting, but they are useful for checking the consistency of the SEDs derived 
from the optical spectra.
The derived stellar masses (Table \ref{tab7}) are relatively low, corresponding 
to that for dwarf galaxies, despite the high brightness due to the active 
ongoing star formation.

The H$\beta$ luminosity $L$(H$\beta$) and SFR were derived from the 
extinction-corrected H$\beta$ flux. Additionally,  
they were also corrected for aperture effects using the relation 
2.512$^{r({\rm ap})-r}$, where $r$ and $r$(ap) are respectively the SDSS 
$r$-band total magnitude and the magnitude 
within the round spectroscopic 3\arcsec\ diameter aperture for J1333$+$6246 and
J1442$-$0209, and within the round spectroscopic 2\arcsec\ diameter aperture
for J1152$+$3400 and J1503$+$3644. According to Table \ref{tab2} the aperture
correction factors are $\sim$ 1.3 -- 1.4. High H$\beta$ luminosities 
(Table \ref{tab8}) are produced by the ionizing radiation of a large number of 
hot massive stars, corresponding to the equivalent numbers of OV7 stars in the 
range (1--3)$\times$10$^5$ assuming the number of ionizing photons per O7V star
of 1$\times$10$^{49}$ s$^{-1}$ \citep{L90}, slightly lower than the value of 
5$\times$10$^5$ for J0925$+$1403. Consequently, the SFRs derived from the 
H$\beta$ luminosity using the relation by \citet{K98}, are also high, 
14 -- 39 M$_\odot$ yr$^{-1}$, and similar to $\sim$ 50 M$_\odot$ yr$^{-1}$
in J0925$+$1403 \citep{I16}.

\begin{figure*}
\hbox{
\includegraphics[angle=-90,width=0.45\linewidth]{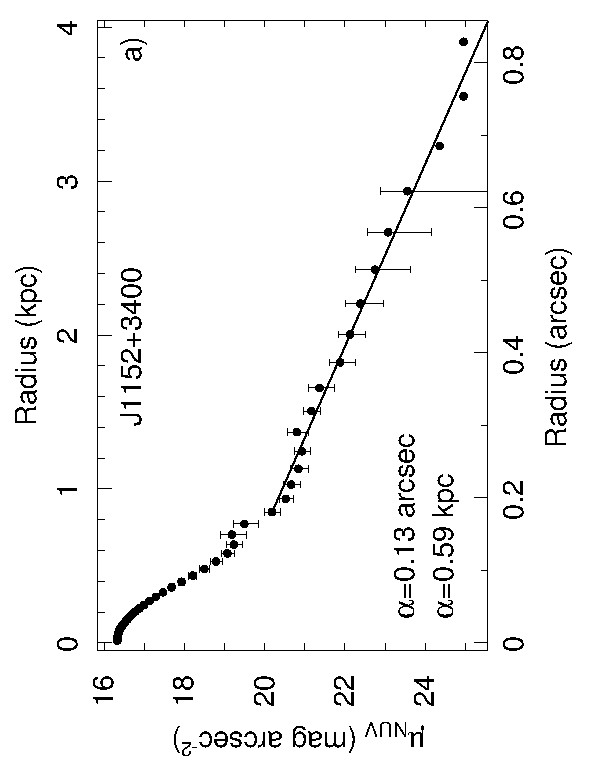}
\includegraphics[angle=-90,width=0.45\linewidth]{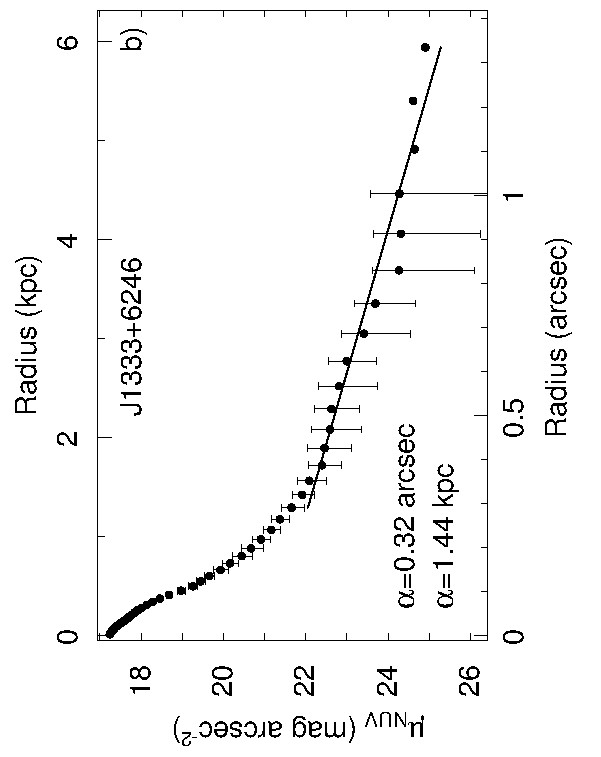}
}
\hbox{
\includegraphics[angle=-90,width=0.45\linewidth]{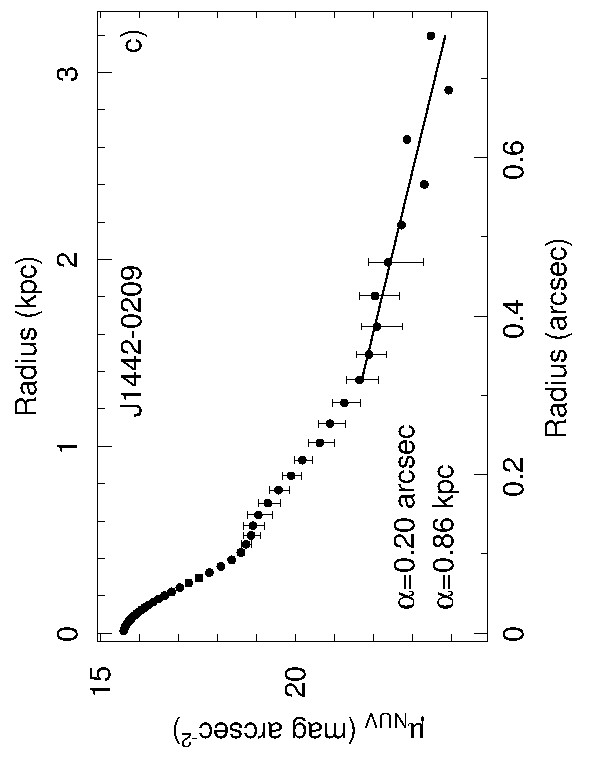}
\includegraphics[angle=-90,width=0.45\linewidth]{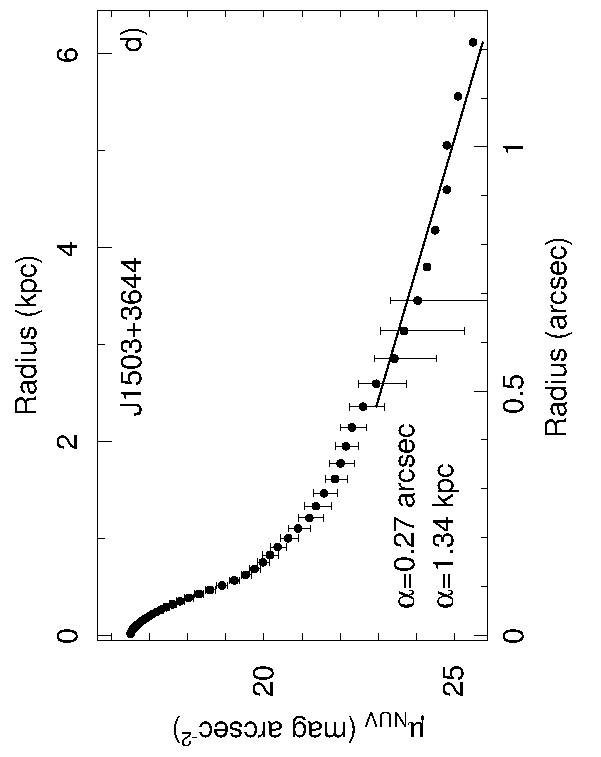}
}
\begin{center}
\includegraphics[angle=-90,width=0.45\linewidth]{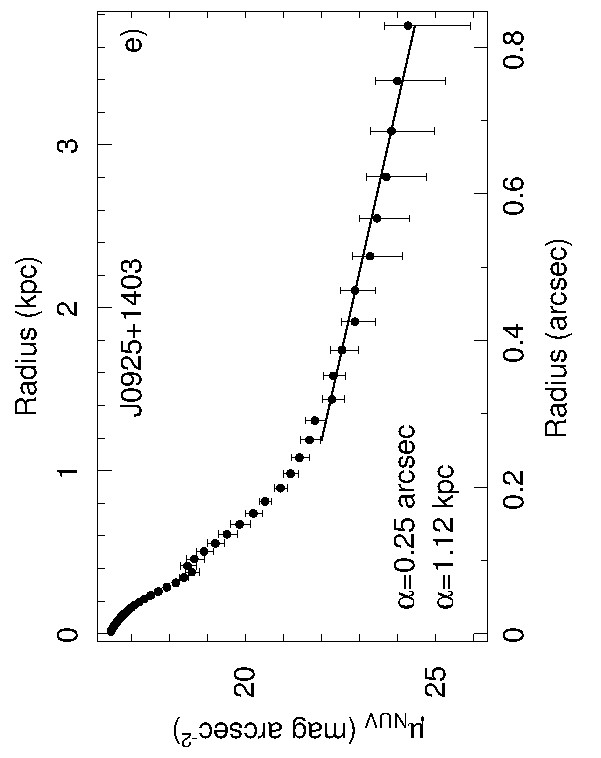}
\end{center}
\caption{Surface brightness profiles of galaxies derived from the COS/NUV 
acquisition images. The linear fits are shown for the range of radii used
for fitting.
\label{fig6}}
\end{figure*}

\begin{figure*}
\includegraphics[angle=-90,width=0.47\linewidth]{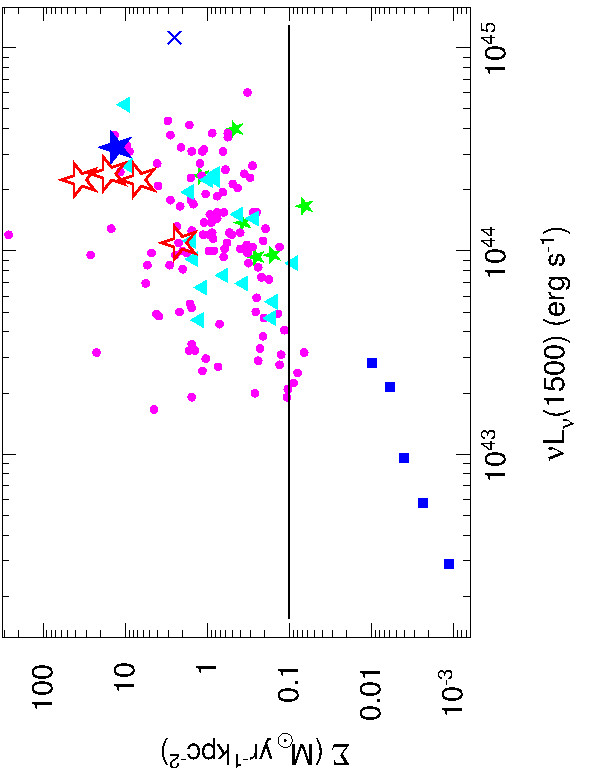}
\caption{The dependence of the mean surface density of star formation rate
$\Sigma$ on the luminosity $\nu L_\nu$(1500). The galaxies from this
paper are shown by large red open stars and the one from \citet{I16} by a large 
blue filled star. These data are corrected for extinction. For comparison, the
average data for SDSS galaxies \citep{BH12} are shown by filled squares,
and LBGs at $z$ = 3, 6,  and 8 \citep{CL16} by small filled stars, circles, 
and triangles. The cross indicates the location of the \citet{B14} galaxy.
The horizontal black line shows the critical $\Sigma$ for outflows \citep{H11}.
\label{fig7}}
\end{figure*}

\begin{figure*}
\hbox{
\includegraphics[angle=-90,width=0.45\linewidth]{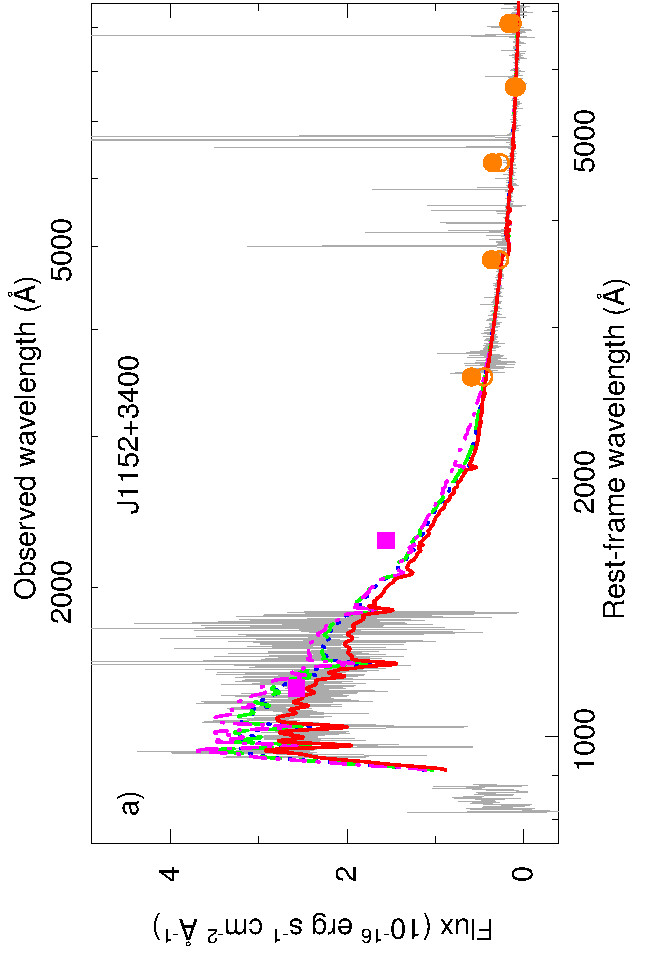}
\hspace{0.5cm}\includegraphics[angle=-90,width=0.45\linewidth]{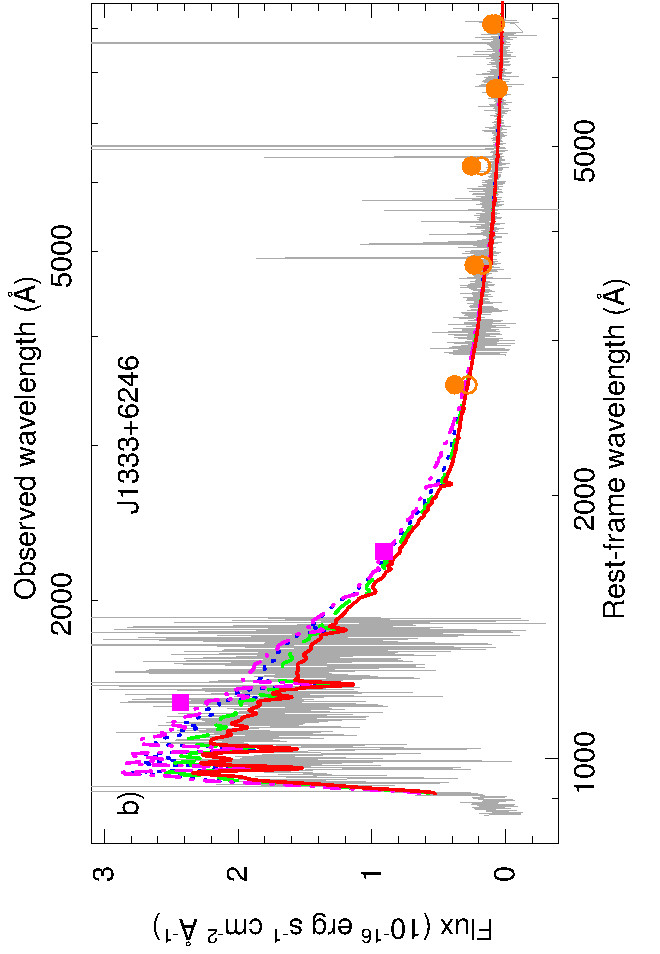}
}
\hbox{
\includegraphics[angle=-90,width=0.45\linewidth]{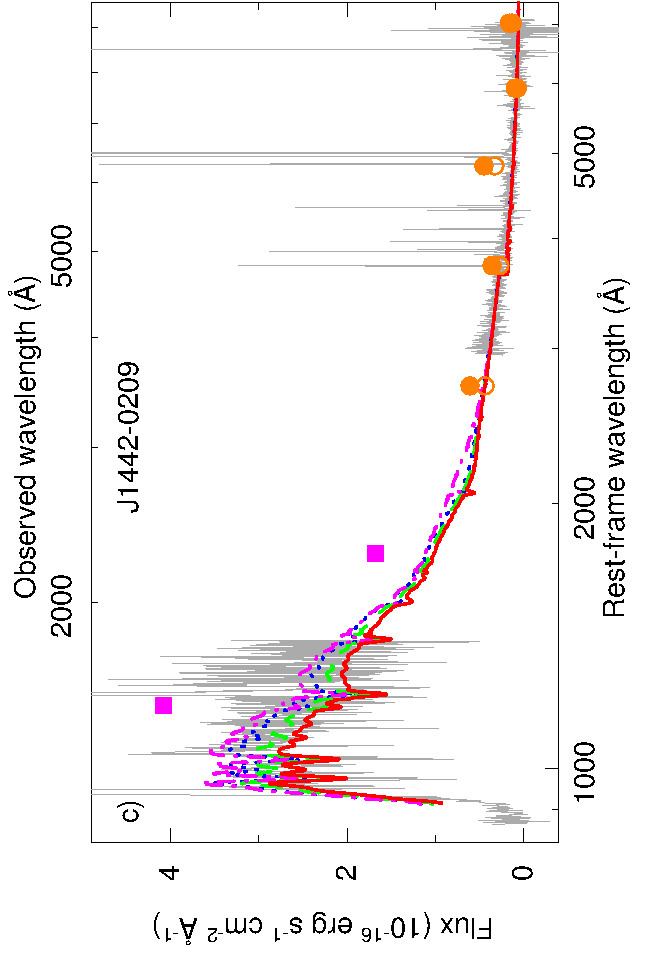}
\hspace{0.5cm}\includegraphics[angle=-90,width=0.45\linewidth]{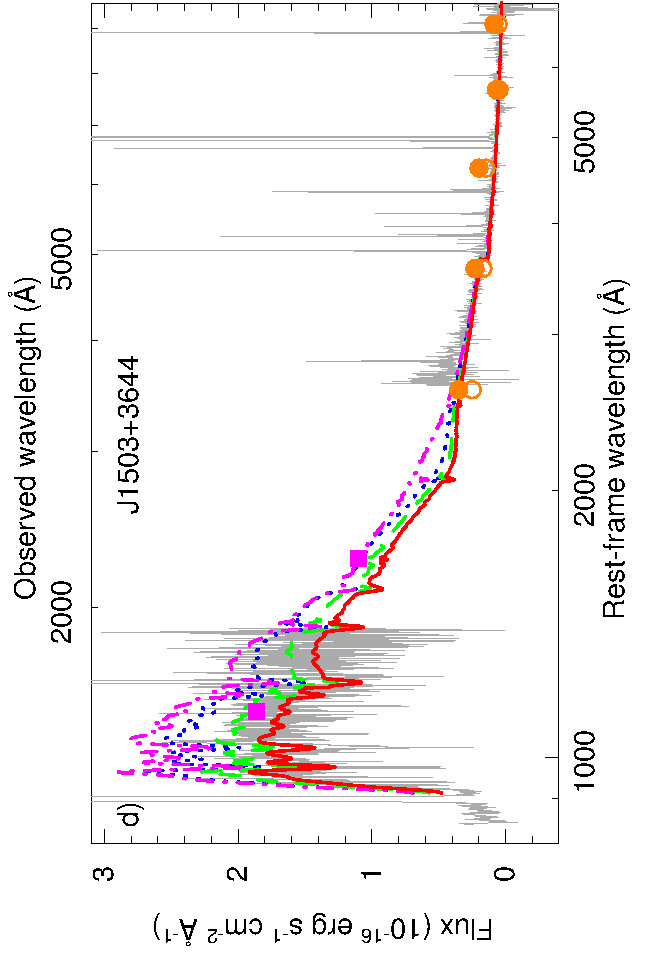}
}
\caption{A comparison of the observed UV and optical spectra, and
photometric data with the modelled SEDs.
The observed spectra are shown by grey lines. The total {\sl GALEX} and SDSS
photometric fluxes are represented by magenta filled squares and orange
filled circles, respectively, while the SDSS photometric fluxes within a 
round spectroscopic aperture of 3\arcsec\ in diameter for J1333$+$6246 and
J1442$-$0209, and of 2\arcsec\ in diameter for J1152$+$3400 and J1503$+$3644 
are shown by open circles. Modelled SEDs, which are reddened by the Milky Way 
extinction with $R(V)_{\rm MW}$ = 3.1 and internal extinction with $R(V)_{\rm int}$
= 3.1, 2.7, and 2.4 adopting the reddening law by \citet{C89} are shown by 
blue dotted, green dashed and red solid lines, respectively, while modelled
SEDs, which reddened by the Milky Way extinction with $R(V)_{\rm MW}$ = 3.1 
adopting the reddening law by \citet{C89} and internal extinction adopting
the reddening law by \citet{C94} are represented by magenta dash-dotted lines.
\label{fig8}}
\end{figure*}

\section{Surface brightness profiles in the NUV range}\label{sec:sbp}

We can use the COS/NUV acquisition images of our galaxies to determine 
their surface brightess (SB) profiles. These images suffer
vignetting at radii greater than 0\farcs5 from the center of the detector. 
In particular, the throughput 
at the radius 0\farcs8 is 0.8 times of that in the center of the
aperture\footnote{http://www.stsci.edu/hst/cos/documents/handbooks/current/ ch06.COS\_Imaging2.html}. Fortunately, radii of our galaxies are small 
(Fig. \ref{fig4}). Therefore, to the first order, we may neglect vignetting
effects. As two NUV frames were obtained during 
acquisition, we combine both images. Then, since transmissions
of {\sl GALEX}/NUV and COS/MIRRORA are
similar\footnote{http://www.stsci.edu/hst/cos/documents/isrs/ISR2010\_10.pdf}$^,$\footnote{http://svo2.cab.inta-csic.es/svo/theory/fps3/index.php?id=GA LEX/GALEX.NUV\&\&mode=browse\&gname=GALEX\&gname2 =GALEX\#filter}, the images were 
approximately reduced to the 
absolute scale using the total {\sl GALEX} NUV magnitudes (Table \ref{tab2}). 
Finally, the routine {\it ellipse} in IRAF\footnote{IRAF is distributed by the 
National Optical Astronomy Observatories, which are operated by the Association
of Universities for Research in Astronomy, Inc., under cooperative agreement 
with the National Science Foundation.}/STSDAS\footnote{STSDAS is product of 
the Space Telescope Science Institute, which is operated by AURA for NASA.} 
was used to produce SB profiles. We derived also the SB profile
of J0925$+$1403 which was not discussed by \citet{I16}.
The SB profiles are shown in Fig. \ref{fig6}a-e. The shape of all profiles is 
very similar, with a sharp SB increase in the central part corresponding to 
the brightest star-forming region in the center of the galaxy, and a linear 
SB decrease (in magnitudes) of the extended LSB component in the outward 
direction.

This linear decrease is characteristic of disc galaxies and can be described
by the relation
\begin{equation}
\mu (r)=\mu (0) + 1.086\times\frac{r}{\alpha},
\end{equation} 
where $\mu$ is the surface brightness, $r$ is the distance from the center, 
and $\alpha$ is the disc scale length of the outer disc.

The derived scale lengths of $\sim$ 0.6 -- 1.4 kpc (Table \ref{tab7}) are larger
by a factor 2--4 than typical scale lengths of blue compact 
dwarf (BCD) galaxies which show very similar morphology \citep[e.g. ][]{P02}. 
They follow however the $\alpha$ -- $M_\star$ relation for the SDSS disc 
galaxies \citep{F10}.
But we note that the scale lengths for BCDs and SDSS disc galaxies are derived 
in the optical or near-infrared broad bands. A direct comparison with the 
scale lengths in the UV range may be somewhat uncertain, as different stellar 
populations may contribute to the light of the LSB component in the UV and 
optical ranges.

\begin{figure*}
\includegraphics[angle=-90,width=0.47\linewidth]{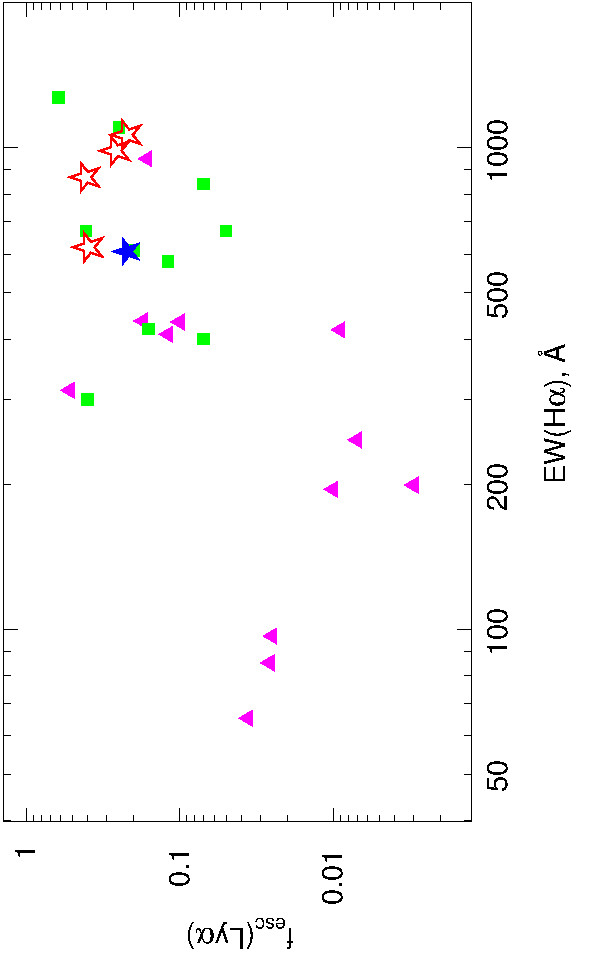}
\caption{The dependence of the Ly$\alpha$ escape fraction
$f_{\rm esc}$(Ly$\alpha$) on the rest-frame equivalent width EW(H$\alpha$) of 
the H$\alpha$ emission line. 
Galaxies from this paper and from \citet {I16}
are represented by red open stars and a blue filled star, respectively. 
For comparison are shown the GPs studied 
by \citet{H15} (green filled squares), and nearby galaxies from the 
Lyman Alpha Reference Sample (LARS) of
\citet{O14} and \citet{H14} (magenta filled triangles).
\label{fig9}}
\end{figure*}

However, the sizes of our galaxies in the UV can directly be compared with those
of galaxies at high redshifts. \citet{CL16} have shown that half-light 
radii of $z$ = 3 -- 8 galaxies are $\sim$ 1 kpc, which are very similar to 
the scale lengths of our galaxies. The compact morphology and high SFR of our 
objects imply high SFR densities $\Sigma$ $\sim$ 2 -- 35 
M$_\odot$ yr$^{-1}$ kpc$^{-2}$ (Table \ref{tab7}, Fig. \ref{fig7}), 
among the highest 
known \citep{LH96}. \citet{Sh16} have argued that galaxies with 
high $\Sigma$s may be efficient at letting ionizing radiation escape.  
High UV luminosities and high SFR densities may
result in outflows which produce channels in the neutral gas, 
allowing ionizing radiation to leak out. However, the data for our galaxies
are not sufficient to verify this possibility.

\section{Reddening law in the UV range}\label{sec:red}

The extinction curve in the optical range is insensitive to variations of 
$R(V)$. On the other hand, the correction of observed fluxes in the UV range 
for extinction critically depends on the adopted reddening law, which
at low metallicities is steeper than the canonical curve with $R(V)$ = 3.1
\citep{B85,GC98,G03}. Therefore, we may expect the reddening law in our 
galaxies to be characterized by $R(V)$ $<$ 3.1. To 
verify that expectation, a comparison of the modelled SEDs with the observed 
photometric and spectroscopic data in the entire UV and optical range is needed.

The observed UV {\sl HST}/COS and optical SDSS spectra are shown in 
Fig. \ref{fig8} by grey lines. Their fluxes are consistent with the SDSS 
photometric fluxes shown by the orange filled circles. As for {\sl GALEX}
UV photometric data (magenta filled squares), they are in agreement with the COS
spectra of J1152$+$3400 and J1503$+$3644 \citep[and J0925$+$1403, ][]{I16}, 
but deviate somewhat in the cases of 
J1442$-$0209 and J1333$+$6246. The deviation is the largest in the FUV band
despite the relatively low quoted {\sl GALEX} magnitude errors of 0.05 mag
and 0.17 mag for J1442$-$0209 and J1333$+$6246, respectively.

To compare with the observed spectra, we reddened the intrinsic modelled 
SEDs in the optical range and their extrapolations in the UV range, adopting 
$R(V)_{\rm MW}$ = 3.1 for the Milky Way extinction, and $R(V)_{\rm int}$ = 3.1 
(blue dotted lines), 2.7 (green dashed lines) and 2.4 (red solid lines) for 
the internal extinction. As before, the Milky Way extinction was applied to
SEDs redshifted to the observed wavelengths and the internal extinction to 
spectra at rest-frame wavelengths.
We adopted the reddening curve by \citet{C89} parameterized by the 
corresponding values of $R(V)_{\rm MW}$ and $R(V)_{\rm int}$, except for 
$\lambda =$ 912 -- 1250\,\AA, where the reddening curve of \citet{M90} with the
respective $R(V)$'s is used. The reddening curve at $\lambda <$ 912\,\AA\
is also provided by \citet{M90}, but it may be somewhat uncertain. 
Fortunately, these
uncertainties do not influence the values of the LyC escape fractions
discussed below, which are determined by the observed to intrinsic flux ratios.
The only reddening correction needed in the UV is for the
observed LyC flux of the Milky Way in the observed wavelength range
1140-1180 \AA, where the reddening law is more reliable.

\begin{figure*}
\hbox{
\includegraphics[angle=-90,width=0.45\linewidth]{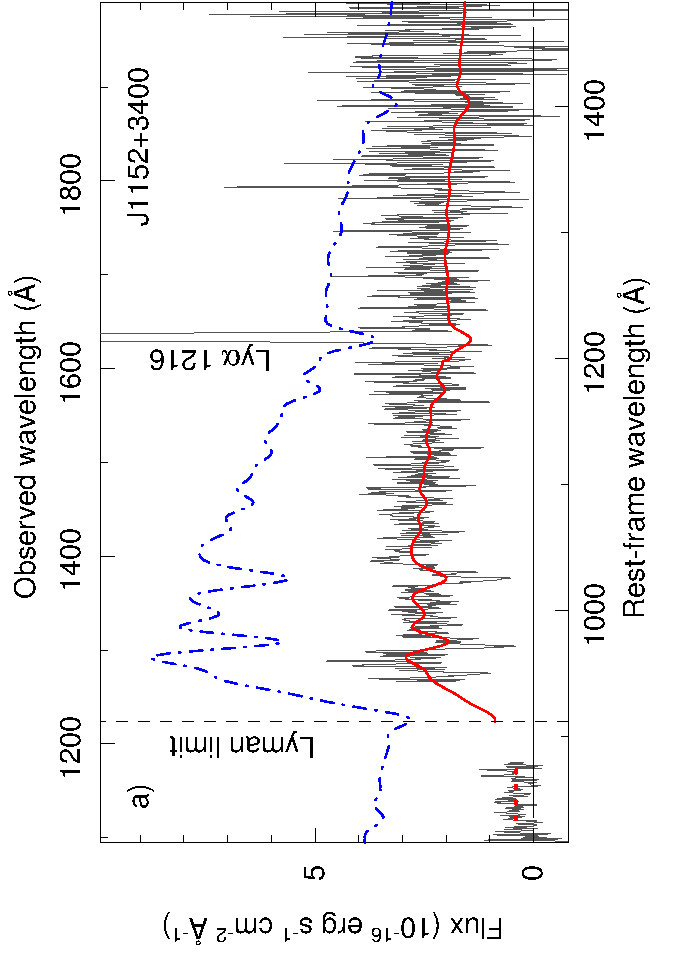}
\hspace{0.5cm}\includegraphics[angle=-90,width=0.45\linewidth]{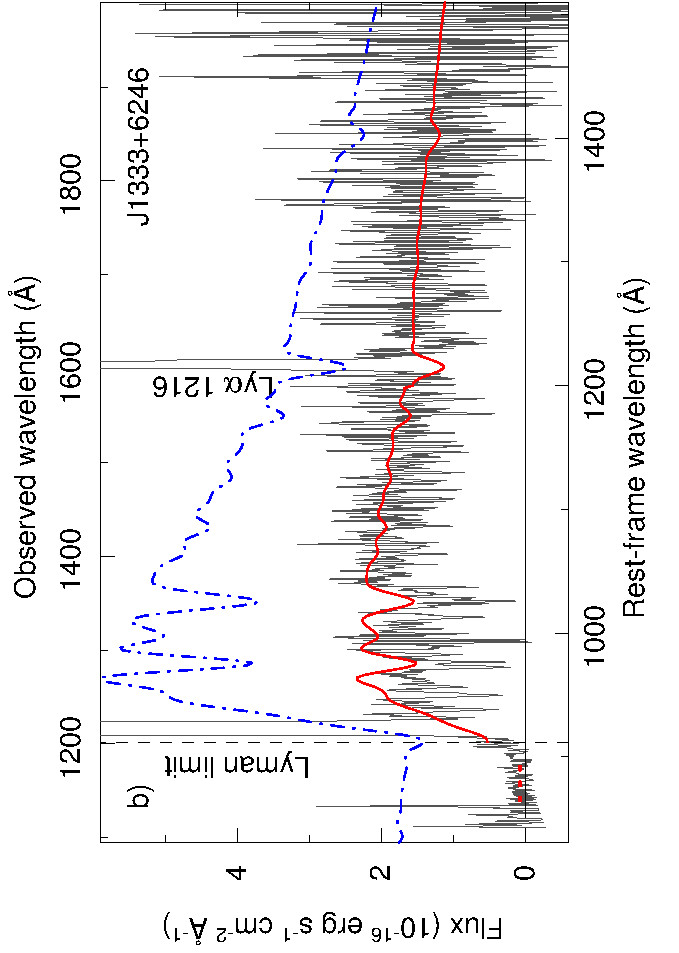}
}
\hbox{
\includegraphics[angle=-90,width=0.45\linewidth]{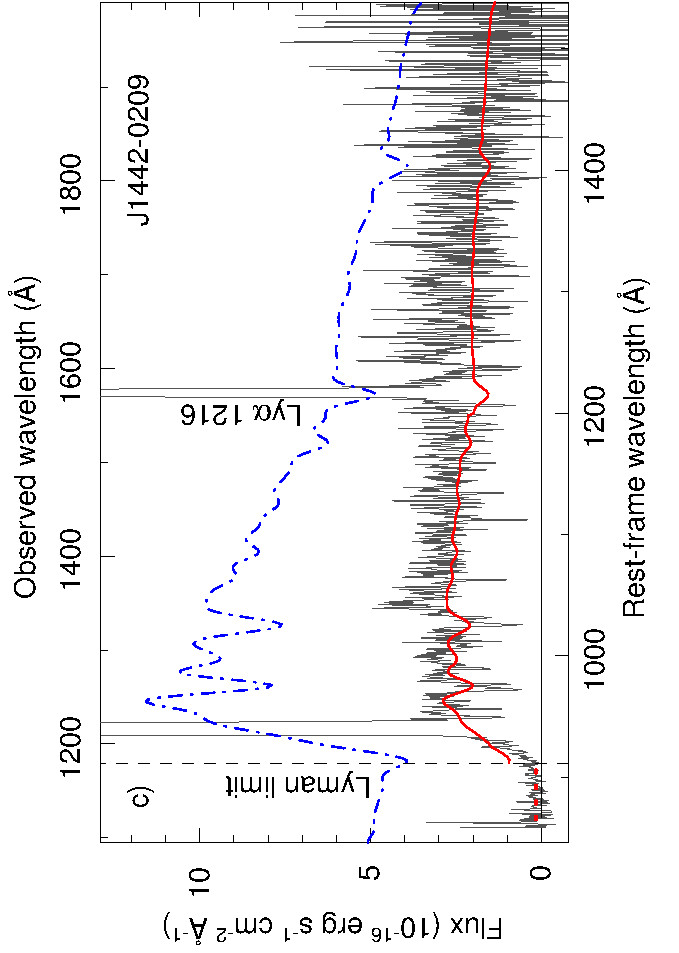}
\hspace{0.5cm}\includegraphics[angle=-90,width=0.45\linewidth]{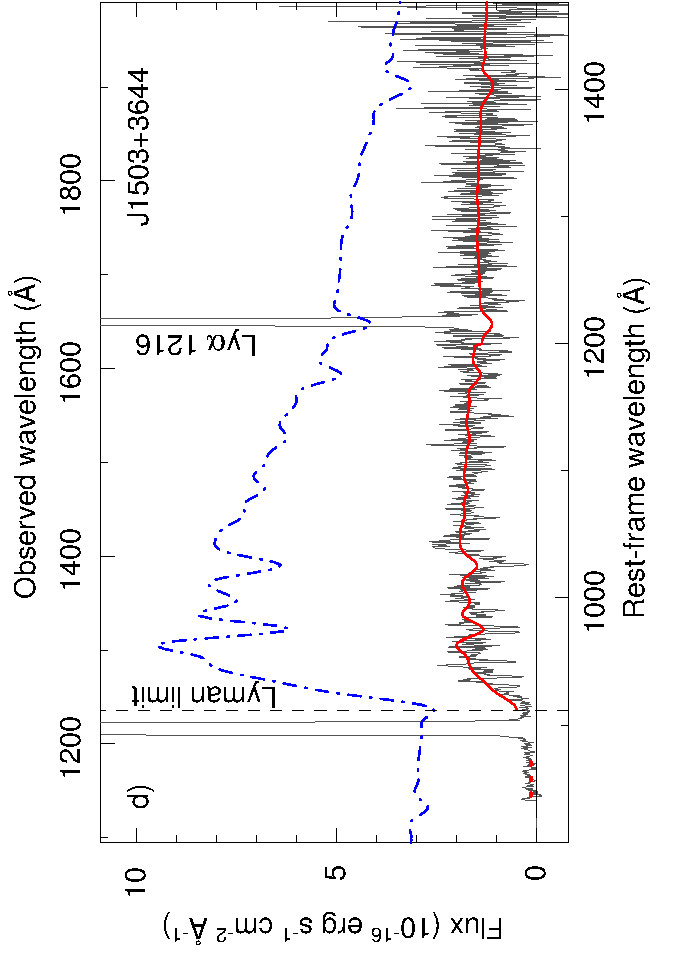}
}
\caption{COS G140L spectra of our sources. On top of the observed
spectra (grey lines) are superposed the modelled SEDs, reddened by 
both the Milky Way and internal extinctions (red solid lines).  
The unreddened SEDs are shown by the blue dash-dotted lines. Zero fluxes are
represented by black solid horizontal lines.
\label{fig10}}
\end{figure*}

\begin{figure*}
\hbox{
\includegraphics[angle=-90,width=0.44\linewidth]{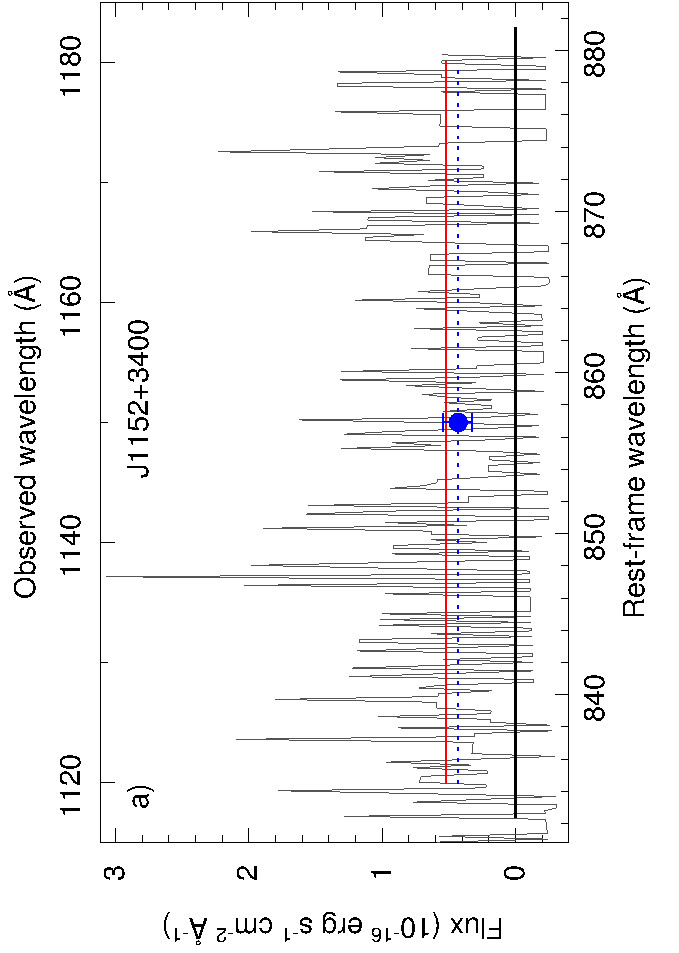}
\hspace{0.5cm}\includegraphics[angle=-90,width=0.44\linewidth]{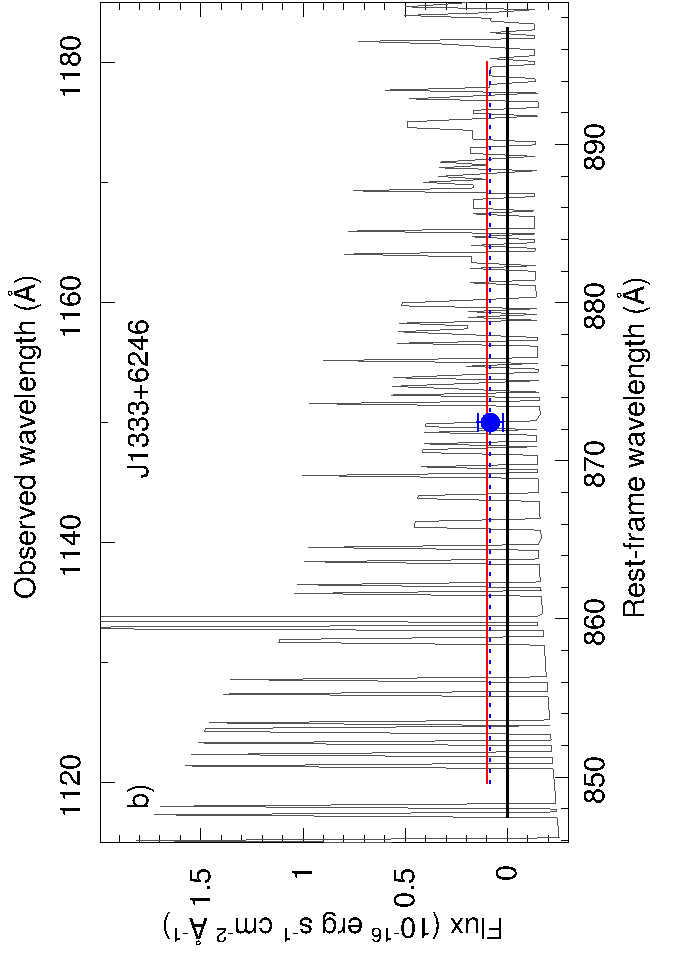}
}
\hbox{
\includegraphics[angle=-90,width=0.44\linewidth]{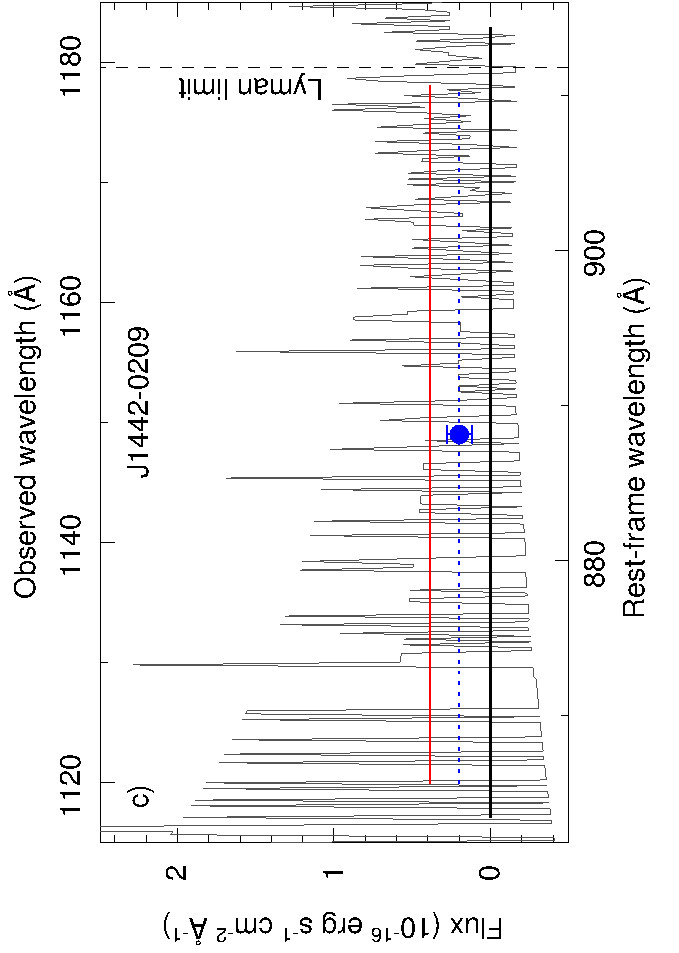}
\hspace{0.5cm}\includegraphics[angle=-90,width=0.44\linewidth]{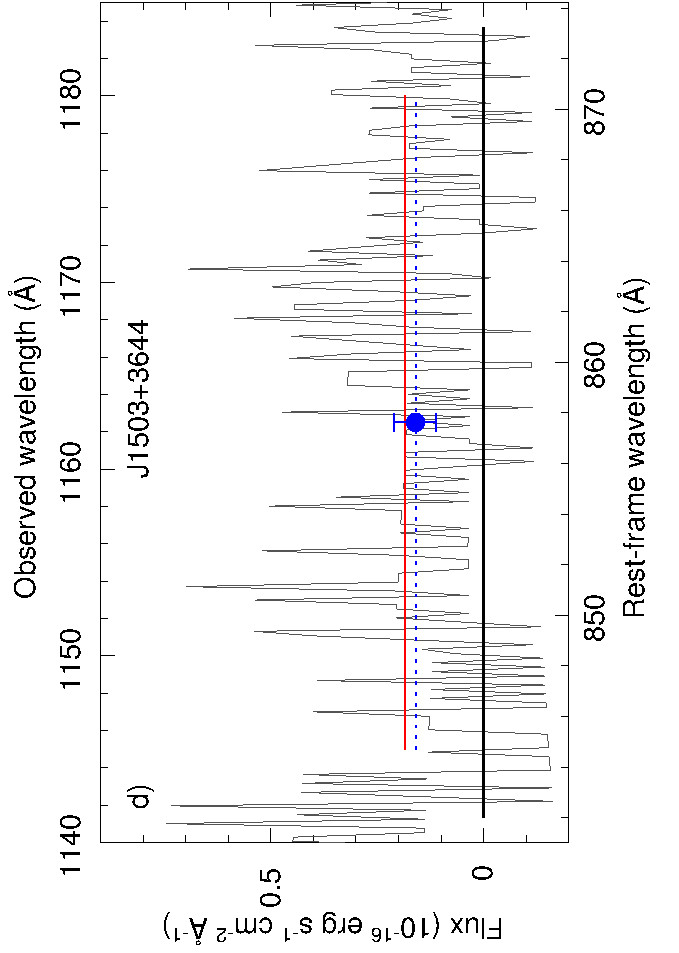}
}
\caption{Expanded COS spectra showing the Lyman continuum. The average values
with the 3$\sigma$ error bars are
shown by blue filled circles. The blue dotted horizontal lines indicate the 
observed mean LyC flux densities and the wavelength ranges used for averaging, 
while the red solid horizontal lines represent the LyC levels after correction 
for the Milky Way extinction. Zero fluxes are also shown by black horizontal
lines.
\label{fig11}}
\end{figure*}

It is seen that the SEDs reddened with $R(V)_{\rm int}$ = 3.1 do not fit 
the observed fluxes in the UV range. A higher extinction coefficient 
is needed in the UV. 
On the other hand, a SED reddened by the same extinctions $A(V)_{\rm MW}$ and
$A(V)_{\rm int}$, but with $R(V)_{\rm int}$ = 2.4, nicely reproduces the observed 
fluxes, with the exception of J1442$-$0209, for which $R(V)_{\rm int}$ = 2.7 is 
more appropriate. 

A similar conclusion has been drawn by \citet{I16} for J0925$+$1403. It appears
that a $R(V)_{\rm int}$ with a value lower than the canonical one of 3.1 is more 
appropriate for dwarf galaxies with low metallicities.

\citet*{C94} and \citet{C00} analyzed the SEDs in the UV range of star-forming 
galaxies and found that star formation occurs in dusty regions. They proposed 
a reddening curve for star-forming galaxies that is less steep than the one 
for the Milky Way and Magellanic Clouds.
If this curve was to be applied to our GP 
galaxies, together with the $A(V)_{\rm int}$ derived from the optical spectrum, 
the reddened modelled SED 
would lie above the observed COS UV spectrum (magenta dash-dotted lines in 
Fig. \ref{fig8}). Thus to fit the observed optical and UV spectra, 
the Calzetti reddening law would require a higher 
extinction in the UV range than that determined from the Balmer decrement.

We note that the galaxies used by \citet{C94} and \citet{C00} 
for the reddening curve determination have properties that are quite
different from ours. In particular, the excitation of 
the H {\sc ii} regions in our galaxies, as 
characterized by the O$_{32}$ ratio, is considerably higher. 
Thus, the \citet{C94} reddening law, determined for dustier and more metal-rich
galaxies, may not be applied to our five galaxies with low metallicities and 
extinctions if the same extinction $A$($V$)$_{\rm int}$ is adopted for the UV and
optical ranges.

\section{Ly$\alpha$ emission}\label{sec:lya}

Strong Ly$\alpha$ $\lambda$1216\,\AA\ emission lines are detected in the 
medium-resolution spectra of all galaxies obtained with the G160M
grating. Their profiles show two peaks with a small separation between the
blue and red peaks, similar to profiles observed in some other GP 
galaxies \citep{JO13,H15,I16,Y16}. 
Correcting for the Milky Way reddening,
we obtain the Ly$\alpha$ flux densities presented in Table \ref{tab8}. 
We find that the extinction-corrected luminosities $L$(Ly$\alpha$)
and rest-frame equivalent widths EW(L$\alpha$) of the Ly$\alpha$ line are among 
the highest known for both low- and high-redshift Ly$\alpha$ emitters 
(Table \ref{tab8}). 

Comparing the extinction-corrected Ly$\alpha$/H$\beta$ flux ratios and the 
case B flux ratio of 23.3 for an electron temperature $T_{\rm e}$ = 10000K and an
electron number density $N_{\rm e}$ = 100 cm$^{-3}$ 
\citep{HS87,SH95}, we find that the Ly$\alpha$ escape fractions are in the 
range $\sim$ 20 -- 40~\%, among the highest known so far for GP galaxies 
\citep[Fig. \ref{fig9},][]{H15,Ha15}.
These high escape fractions are consistent with a low H~{\sc i} column density,
and the very young age of the starbursts in our galaxies, as evidenced by 
their high EW(H$\beta$)s.
A similar high value of $f_{\rm esc}$(Ly$\alpha$) was obtained for J0925$+$1403
\citep{I16}. 

The double-peaked Ly$\alpha$ line profile of J0925$+$1403 was briefly 
discussed in \cite{I16}.
The Ly$\alpha$ line profiles for the entire sample will be shown and analyzed 
in a separate paper (Verhamme et al., in preparation).

  \begin{table*}
  \caption{Parameters for the H$\beta$ and Ly$\alpha$ emission 
lines \label{tab8}}
  \begin{tabular}{lccccccrcrccccc} \hline
    &\multicolumn{3}{c}{}&&\multicolumn{3}{c}{H$\beta$}&&\multicolumn{3}{c}{Ly$\alpha$}&& \\
Name&$A(V)_{\rm MW}^{\rm a}$&$A(V)_{\rm int}^{\rm a}$&$A$(Ly$\alpha$)$^{\rm a}_{\rm MW}$&&$F$$^{\rm b}$&EW$^{\rm c}$&$L$$^{\rm d}$
&&$F$$^{\rm b}$&EW$^{\rm c}$&$L$$^{\rm d}$&Ly$\alpha$/H$\beta$$^{\rm e}$&$f_{\rm esc}$(Ly$\alpha$)$^{\rm f}$ \\ \hline
J1152$+$3400&0.054&0.185&0.140&&23.8&148&1.76&& 211.3& 79&10.88& 6.2&0.26\\
J1333$+$6246&0.052&0.152&0.135&& ~\,9.8&147&0.62&& 137.1& 75& ~\,5.56& 9.0&0.39\\
J1442$-$0209&0.148&0.185&0.389&&29.8&241&1.60&& 358.1&129~\,&15.25& 9.5&0.41\\
J1503$+$3644&0.041&0.283&0.106&&19.2&219&1.70&& 178.3& 98& ~\,9.15& 5.4&0.22\\
\hline
  \end{tabular}


\hbox{$^{\rm a}$Extinction in mags. Data for J0925$+$1403 are presented in \citet{I16}.}

\hbox{$^{\rm b}$Observed flux density in 10$^{-16}$ erg s$^{-1}$ cm$^{-2}$. Data for J0925$+$1403 are presented in \citet{I16}.}

\hbox{$^{\rm c}$Rest-frame equivalent width in \AA. EW(Ly$\alpha$) for J0925$+$1403 is
83\AA.}

\hbox{$^{\rm d}$Luminosity in 10$^{42}$ erg s$^{-1}$. $L$(H$\beta$) is corrected for both
the Milky Way and internal extinction, $L$(Ly$\alpha$) is corrected 
}
\hbox{only for the Milky Way extinction.}

\hbox{$^{\rm e}$Luminosity ratio. $L$(Ly$\alpha$)/$L$(H$\beta$) for J0925$+$1403 is 5.5.}

\hbox{$^{\rm f}$$f_{\rm esc}$(Ly$\alpha$) for J0925$+$1403 is 0.22.}
  \end{table*}

\begin{figure*}
\hbox{
\includegraphics[angle=-90,width=0.44\linewidth]{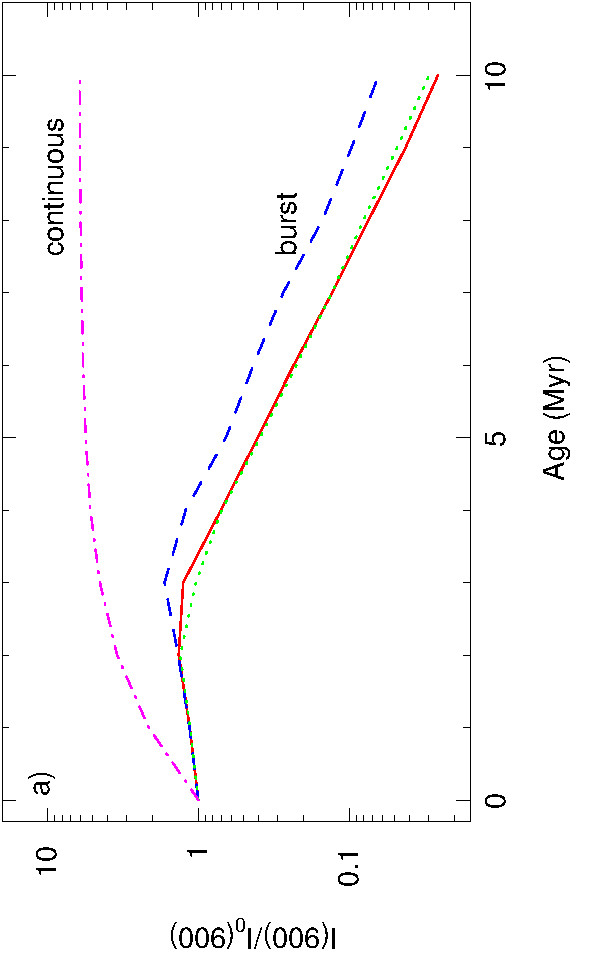}
\hspace{0.5cm}\includegraphics[angle=-90,width=0.44\linewidth]{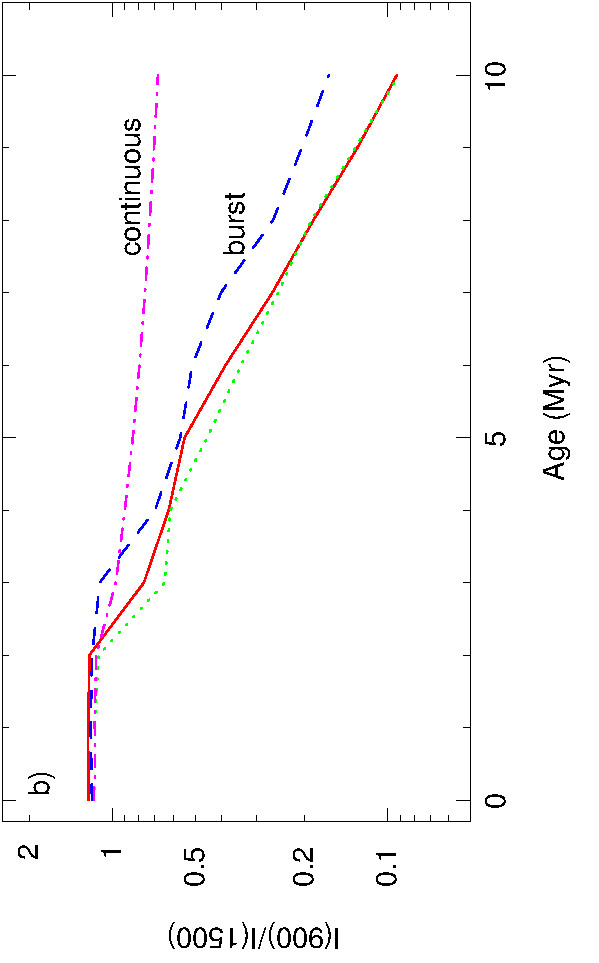}
}
\vspace{0.3cm}
\hbox{
\includegraphics[angle=-90,width=0.44\linewidth]{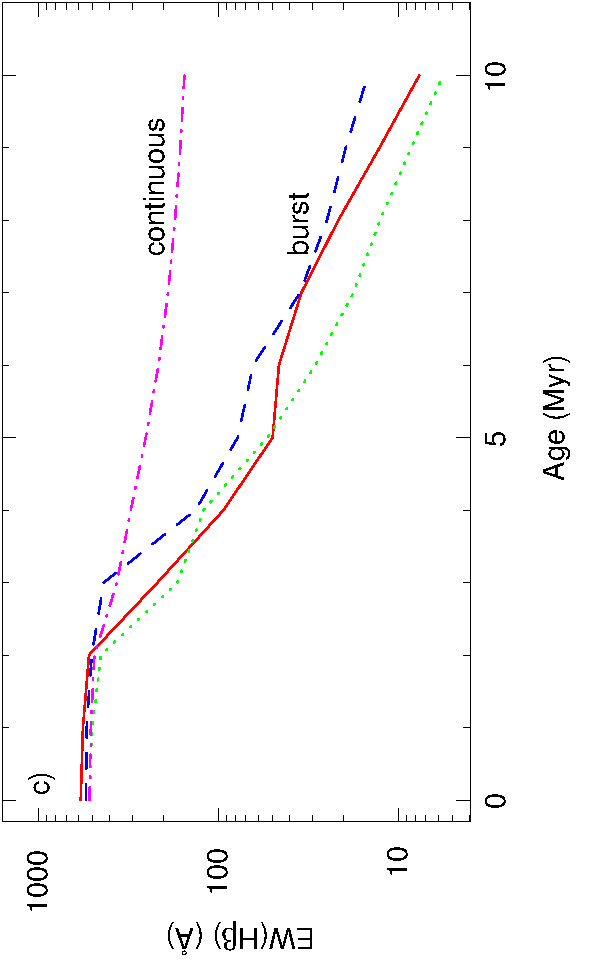}
\hspace{0.5cm}\includegraphics[angle=-90,width=0.44\linewidth]{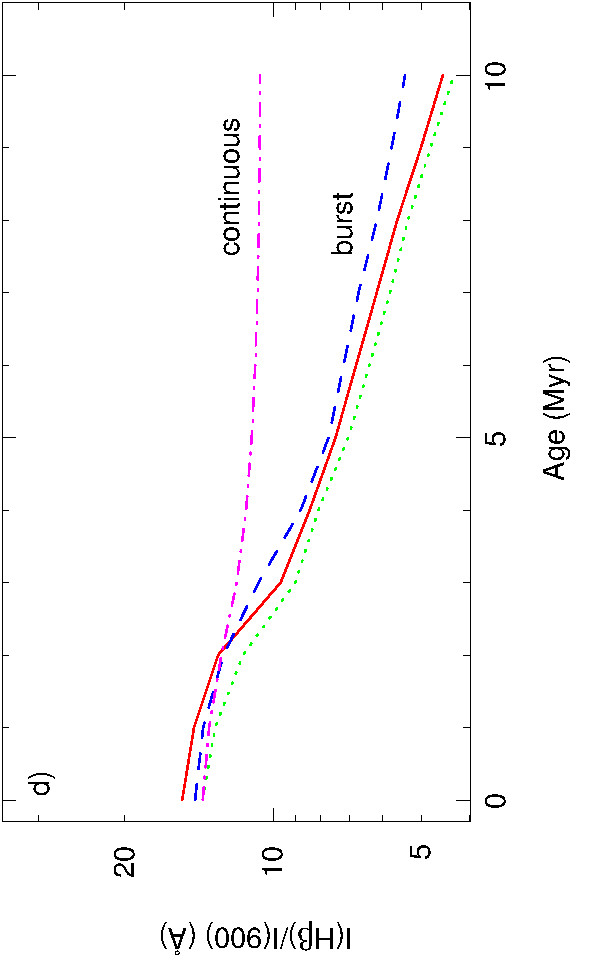}
}
\caption{Dependence on the starburst age of: a) the flux at 900\AA\ 
normalized to the initial flux, b) the flux ratio at wavelengths
900\AA\ and 1500\AA\ (in $f_\lambda$), c) the H$\beta$ equivalent width EW(H$\beta$), and d) 
the H$\beta$-to-900\AA\ flux ratio. Red solid lines and blue dashed lines are 
model predictions for instantaneous bursts with stars rotating with velocities
0 km s$^{-1}$ and 400 km s$^{-1}$, respectively \citep{L14}, green dotted lines
are model predictions from \citet{L99}. Magenta dash-dotted lines are 
relations for continuous star formation with a constant SFR and models
by \citet{L99}. For all calculations, a \citet{S55} IMF
and stellar atmosphere models by \citet{L97} and \citet{S92} are adopted.
\label{fig12}}
\end{figure*}

\begin{figure*}
\includegraphics[angle=-90,width=0.47\linewidth]{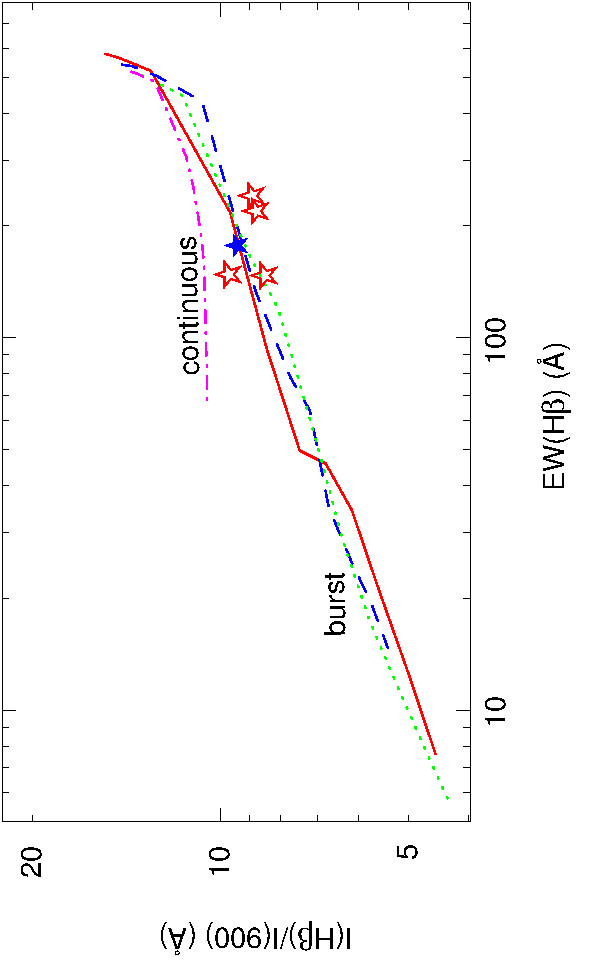}
\caption{The dependence of the H$\beta$-to-900\AA\ flux ratio on the
H$\beta$ equivalent width EW(H$\beta$) for instantaneous burst and continuous 
SFR models with negligible Lyman continuum 
escape. The meaning of lines is the same as in Fig. \ref{fig12}. The location of
LyC leaking galaxies from \citet{I16} and this paper are shown by a
blue filled star and red open stars, respectively.
\label{fig13}}
\end{figure*}

\section{Lyman continuum detection and the inferred escape fraction of ionizing photons}\label{sec:lyc}

We now examine the COS spectra obtained with the G140L grating,
primarily to examine the Lyman continuum of our sources, and inferences from 
these data.

\subsection{Detection of Lyman continuum leakage}

The observed G140L spectra are shown in Fig. \ref{fig10} together with the 
predicted intrinsic SEDs obtained from fitting the observed optical SEDs
(cf.\ below). The strong line at the observed wavelength of 
1216\AA, present in all objects except J1152+3400, is the residual of the
geocoronal Ly$\alpha$ emission (not labelled), while the second brightest line
labelled ``Ly$\alpha$ 1216'' is the Ly$\alpha$ line of the galaxy. 
The most striking finding is the detection of a non-zero flux in the Lyman 
continuum in all our sources. This is illustrated more clearly in 
Fig. \ref{fig11}, which shows a blow-up of the LyC spectral region
for all four galaxies. Similarly to J0925$+$1403 \citep{I16}
the important feature is that the Lyman continuum flux density for the 
rest-frame $\lambda$ $<$ 912\AA\ is not zero, but positive, when averaged over 
the spectral range indicated by blue dotted and red solid horizontal lines. 
The lower cut on the wavelength range is typically due to declining 
sensitivity at the lowest wavelengths, but for J1503$+$3644 there is residual 
geocoronal N~{\sc i} 1134\AA\ emission that needs to be avoided. The limit
$\lambda < 1180$\AA\ allows to avoid scattered geocoronal 
Ly$\alpha$ and/or wavelengths redward of 912\AA\ in the rest frame of the 
galaxy. The floor at negative flux values is due to background subtraction 
(sometimes negative flux) and declining COS sensitivity. At the blue end one 
Poisson count corresponds to a larger ``flux quantum'', such that the negative 
floor is curved downward.

With fluxes in the range $(8-43) \times 10^{-18}$ erg~s$^{-1}$~cm$^{-2}$~\AA$^{-1}$
the Lyman continuum is detected at the $>$ 10$\sigma$ level in the J1152$+$3400 
and J1503$+$3644 spectra, at the $\sim$ 8$\sigma$ level in the J1442$-$0209 
spectrum, and at the $\sim$ 5$\sigma$ level in the J1333$+$6246 spectrum
(Table \ref{tab9}). The observed mean
levels of the LyC are indicated in Fig. \ref{fig11} by blue dotted horizontal 
lines and blue filled circles. For comparison, the LyC detection of our first 
target, J0925$+$1403, is at the 11.8 $\sigma$ level \citep{I16}.

The observed LyC emission should be corrected for extinction from the Milky Way 
before the determination of the LyC escape fraction. The average LyC flux 
densities corrected for the Milky Way extinction are shown in Fig. \ref{fig11} 
by the red solid horizontal lines. The observed and corrected flux densities 
are reported in Table \ref{tab9}.

  \begin{table*}
  \caption{LyC escape fraction \label{tab9}}
\begin{tabular}{lccccccclcc} \hline
 &\multicolumn{5}{c}{Monochromatic flux density$^{\rm a}$}& \\ 
Name&$I_{\rm mod}$$^{\rm b}$&$I_{\rm mod}$$^{\rm b}$&$I_{\rm obs}$$^{\rm c}$&$I_{\rm obs}$$^{\rm c}$&$I_{\rm esc}$$^{\rm d}$&22$\mu$m/&$f_{\rm esc}^{\rm rel}$$^{\,\rm f}$&\multicolumn{1}{c}{$f_{\rm esc}$$^{\rm g}$}&$f_{\rm esc}$$^{\rm h}$&$f_{\rm esc}$$^{\rm i}$ \\
    &(900\AA)&(1500\AA)&(900\AA)&(1500\AA)&(900\AA)&1500\AA$^{\rm e}$&(LyC)&\multicolumn{1}{c}{(LyC)}&(LyC)&(LyC) \\
\hline
J0925$+$1403&4.40&5.84&0.235$^{+0.021}_{-0.020}$&1.55&0.343$^{+0.038}_{-0.037}$&1.08    &0.236&0.072$\pm$0.008$^{\rm j}$&0.064&0.046 \\
J1152$+$3400&3.30&3.14&0.428$^{+0.037}_{-0.034}$&1.36&0.502$^{+0.043}_{-0.040}$&  ...   &0.335&0.132$\pm$0.011&0.132&0.116 \\
J1333$+$6246&1.52&2.05&0.083$^{+0.023}_{-0.022}$&0.75&0.090$^{+0.025}_{-0.024}$&  ...   &0.149&0.056$\pm$0.015&0.064&0.055 \\
J1442$-$0209&4.59&3.99&0.198$^{+0.027}_{-0.028}$&1.50&0.367$^{+0.050}_{-0.052}$&1.44    &0.154&0.074$\pm$0.010&0.090&0.032 \\
J1503$+$3644&3.19&3.24&0.160$^{+0.018}_{-0.017}$&1.80&0.195$^{+0.021}_{-0.019}$&1.54    &0.090&0.058$\pm$0.006&0.069&0.050 \\ \hline
  \end{tabular}

\hbox{$^{\rm a}$in 10$^{-16}$ erg s$^{-1}$cm$^{-2}$\AA$^{-1}$.}

\hbox{$^{\rm b}$flux density derived from the modelled SED.}

\hbox{$^{\rm c}$observed flux density.}

\hbox{$^{\rm d}$observed flux density which is corrected for the Milky Way extinction.}

\hbox{$^{\rm e}$$\nu L_\nu$(22$\mu$m)/[$\nu L_{\nu,\rm mod}$(1500) -- $\nu L_{\nu,\rm obs}$(1500)].}

\hbox{$^{\rm f}$$I_{\rm mod}$(1500)/$I_{\rm obs}$(1500)$\times$$I_{\rm obs}$(900)/$I_{\rm mod}$(900).}

\hbox{$^{\rm g}$$I_{\rm esc}$(900)/[$I_{\rm mod}$(900)+$I_{\rm esc}$(900)], where $I_{\rm mod}$(900) is derived from SED (first method).}

\hbox{$^{\rm h}$$I_{\rm esc}$(900)/[$I_{\rm mod}$(900)+$I_{\rm esc}$(900)], where $I_{\rm mod}$(900) is derived from Eq. \ref{eq:i900} (second method).}

\hbox{$^{\rm i}$Derived from Eq. \ref{eq:esc} \citep[see also ][]{L13}.}

\hbox{$^{\rm j}$This value is somewhat lower than the value $f_{\rm esc}$ = $I_{\rm esc}$(900)/$I_{\rm mod}$(900) = 0.078 derived by \citet{I16}.}
  \end{table*}

\subsection{The Lyman continuum escape fraction}

To derive the absolute escape fraction of Lyman continuum photons, \fesclyc,
we need to use the modelled flux $I(900)$ of the Lyman continuum 
emission at 900 \AA,
as produced by the massive stars, and compare it to the observed level
(after correction for extinction in the Milky Way).

Two methods may be proposed to determine $I(900)$. 
The first method is based on ``global" SED fits to the observed spectra 
and/or photometry, using evolutionary synthesis models, which then predict 
the intrinsic UV emission, both long- and short-ward of the Lyman limit.
The second method relies on the fact that the intensities of hydrogen 
recombination lines are proportional to the number of 
ionizing photons emitted per unit time, $N$(Lyc). Both methods use 
the extinction-corrected flux $I_{\rm cor}$(H$\beta$) and rest-frame equivalent
width EW(H$\beta$) of the H$\beta$ emission line which are lower in LyC 
leaking galaxies because escaping ionizing radiation is not processed to
produce nebular continuum and recombination line emission. 
Therefore, both methods should give accurate 
results if $f_{\rm esc}$(LyC) is much lower than unity.
The advantage of the first
method is that it takes into account the emission of the young and old 
stellar populations and thus potentially allows to derive more accurately the 
starburst age.
On the other hand, the second method is simpler because no modelling is needed.
However, it requires assumptions concerning the star formation history. 
 
Concerning the first method, the modelled UV flux densities 
$I_{\rm mod}$($\lambda$) of young clusters with the ages
obtained from SED fitting in the optical range are shown 
by the blue dash-dotted lines in Fig. \ref{fig10}.
The reddened modelled UV SEDs are represented by the red 
solid lines. These SEDs are the same as the SEDs shown by red solid lines 
in Fig. \ref{fig8}.
As discussed above (Sects.\ \ref{sec:global} and \ref{sec:red}), 
we adopt for the reddening $R(V)_{\rm MW}$ = 3.1, $A(V)_{\rm MW}$ from the NED,
$R(V)_{\rm int}$ = 2.4, $A(V)_{\rm int}$ as derived from the Balmer decrement
after correction of the SDSS spectra for the Milky Way extinction.
It is seen that the reddened SEDs reproduce very well 
the observed spectra (grey lines) for rest-frame wavelengths $>$ 912\AA.

The predicted flux densities $I_{\rm mod}(900)$ at 900 \AA\ obtained from the best
SED fits are listed in Table \ref{tab9}. To derive the escape fraction
\fesclyc\ we need to derive the total flux density produced by massive stars
\begin{equation}
I(900) = I_{\rm mod}(900) + I_{\rm esc}(900), \label{eq:Itot}
\end{equation}
where $I_{\rm mod}(900)$ is the flux density of radiation producing the
H$\beta$ emission and $I_{\rm esc}(900)$ is the flux density of escaping 
radiation (Table \ref{tab9}). Then the escape fraction is obtained from
\begin{equation}
f_{\rm esc}({\rm LyC}) =\frac{I_{\rm esc}(900)}{I(900)}=
\frac{I_{\rm esc}(900)}{I_{\rm mod}(900)+I_{\rm esc}(900)}. \label{eq:fesc}
\end{equation}

Using Eq. \ref{eq:fesc} we derive escape fractions  \fesclyc\ in the range 
between $\sim$ 6 \% for J1333$+$6246 and J1503$+$3644, and 13 \% for 
J1152$+$3400.

For the second method we need to consider the relation between 
$N$(LyC), which is directly related to the extinction-corrected flux
density $I$(H$\beta$), and the flux density $I(900)$ at the reference wavelength
representative of our LyC observations adopting a negligible LyC escape
fraction. Since $N$(LyC) $\propto \int_{0}^{912} I(\lambda) / \lambda d\lambda$,
and the shape of the ionizing spectrum $I(\lambda)$ depends e.g.\ on starburst 
age, the ratio $I$(H$\beta$)/$I(900)$ is not constant. 
However, it can be fairly accurately constrained, as we will now show.
In Fig. \ref{fig12} we examine how well the H$\beta$ intensity traces the LyC 
flux $I(900)$ and how this depends on different IMFs, stellar 
atmospheres, 
and tracks. For a 1/10 solar metallicity, we plot how different quantities and
their ratios which depend on the LyC flux, evolve with age of the burst 
stellar population. These relations were calculated with
{\em Starburst99} models \citep{L99,L14} adopting a \citet{S55} IMF,
a combination of stellar atmosphere models by \citet{L97} and \citet{S92},
and various Geneva evolutionary tracks for non-rotating and rotating stars.
The LyC flux $I$(900) produced during the first 3 Myr of the burst is 
nearly constant because the most massive stars with masses $\sim$ 100 M$_\odot$ 
are still on the main sequence and
are producing ionizing radiation (Fig. \ref{fig12}a). Later, between 3 and 10 
Myr, the LyC flux rapidly decreases, by a factor of $\sim$ 30. Similarly,
$I$(900)/$I$(1500) and EW(H$\beta$) decrease by a factor $\sim$10 and $\sim$30, 
respectively, during the same time period (Fig. \ref{fig12}b and 
\ref{fig12}c). On the other hand, the $I$(H$\beta$)/$I$(900) ratio diminishes
only by a factor $\sim 3$ (Fig. \ref{fig12}d), and is consistent
with the \citet{I16} relation for younger ages. 
The decrease of $I$(H$\beta$)/$I$(900) is explained by the evolution of the 
shape of the ionizing spectrum with time. By contrast, the relations in 
Fig. \ref{fig12} for continuous star formation
with a constant SFR converge asymptotically toward constant values after an 
age of $\sim$ 5 Myr.

To determine the flux density $I$(900) = $I_{\rm mod}$(900) + $I_{\rm esc}$(900), 
it is more convenient to use relations 
between the $I$(H$\beta$)/$I$(900) ratio and EW(H$\beta$), since both
the H$\beta$ flux and the H$\beta$ equivalent width
can directly be derived from observations. These relations for
instantaneous bursts and different sets of stellar atmosphere models are 
shown in Fig.~\ref{fig13} (labelled ``burst''), and can be 
approximated as
\begin{equation}
\frac{I_{\rm cor}({\rm H}\beta)}{I_{\rm mod}(900)} \approx \frac{I({\rm H}\beta)}{I(900)} = 
2.99\times{\rm EW}({\rm H}\beta)^{0.228} \,\,\,\,{\rm \AA} \label{eq:i900}
\end{equation}
for a negligible LyC escape fraction, where EW(H$\beta$) is in \AA.
As can be seen, this relation is fairly weakly 
model-dependent, which allows us to determine quite accurately the flux
$I_{\rm mod}(900)$ (within $\leq$ 10\%) adopting the rest-frame equivalent 
width EW(H$\beta$) from Table \ref{tab8} and the extinction-corrected H$\beta$ 
flux $I_{\rm cor}$(H$\beta$) from Table \ref{tab5}. The characteristic value 
$I_{\rm cor}$(H$\beta$)/$I_{\rm mod}$(900) for our galaxies (shown by stars) is $\sim$8--10.
For continuous star formation with a constant SFR, the ratio attains
the asymptotic value of $\sim$ 11 (labelled 
``continuous'' in Fig. \ref{fig13}).
We note that, in general, EW(H$\beta$) in galaxies is not only determined by 
the recent starburst, but also by the radiation of older stellar populations 
in the optical range. But for our galaxies, EW(H$\beta$) is mainly set by 
the young stellar population. Otherwise, the intrinsic EW(H$\beta$) to be
compared to the models in Fig. \ref{fig13} would be even higher.
However, our adopted value of $I_{\rm cor}$(H$\beta$)/$I_{\rm mod}$(900) would not change much.

The escape fraction with the second method is derived from Eq. \ref{eq:fesc}
(similar to the first method), but with $I_{\rm mod}$(900) obtained
from Eq. \ref{eq:i900} and adopting a Salpeter IMF \citep{S55}, Geneva 
evolutionary tracks \citep{M94} of non-rotating stars and a combination of 
stellar atmosphere models \citep{L97,S92}. It ranges from 
$\sim$ 6 \% to $\sim$ 13 \% (Table \ref{tab9}) with the errors, which are 
similar to those given in Table \ref{tab9} for method 1.
 
These values agree within the errors with the
values from the first method based on SED fitting (cf.\ above).
We decided to adopt the $f_{\rm esc}$ derived by method 1, as the values
derived by method 2, not based on modelling of the SEDs, are more
approximate.

  The $f_{\rm esc}$(LyC) values, derived by either method,
are higher than the $f_{\rm esc}$(LyC) of the other four
low-redshift galaxies with known LyC leakage.

In the literature, the LyC escape fraction is often estimated from 
the equation \citep{L13}
\begin{equation}
f_{\rm esc}({\rm LyC})=f_{\rm esc}^{\rm rel}({\rm LyC})\times 10^{-0.4\times A(1500)},  
\label{eq:esc}
\end{equation}
where 
\begin{equation}
f_{\rm esc}^{\rm rel}({\rm LyC})
=\frac{[I(1500)/I(900)]_{\rm mod}}{[I(1500)/I(900)]_{\rm obs}}, \label{eq:relesc}
\end{equation}
$I(900)$ and $I(1500)$ are the flux densities at rest-frame wavelengths 
900\AA\ and 1500\AA, respectively, and the subscripts ``mod'' and ``obs'' 
denote modelled and observed flux densities. The values of the relative 
escape fraction $f_{\rm esc}^{\rm rel}$(LyC) for our objects, derived from SED
fitting and listed in Table \ref{tab9}, are between 9 and 33 \%. The use of 
Eq. \ref{eq:esc} would result in an underestimate of $f_{\rm esc}$(LyC) for
two reasons: 1) Eq. \ref{eq:esc} does not take into account the fact that the 
extinction at 900\AA\ though uncertain is expected to be higher than at 
1500\AA\ \citep{M90}; and 2) the observed flux density $I_{\rm obs}$(900) is not
corrected for foreground extinction from the Milky Way as it should.
Indeed, using $I_{\rm mod}$, $I_{\rm obs}$ and 10$^{-0.4\times A(1500)}$ $\equiv$
$I_{\rm obs}$(1500)/$I_{\rm mod}$(1500) we find that, 
depending on the object, the $f_{\rm esc}$(LyC) derived from Eq. \ref{eq:esc} 
is up to 55 \% lower than the value obtained from SED fitting 
(last column in Table \ref{tab9}).

In the above determination of $f_{\rm esc}$(LyC) by different methods, 
the UV attenuation is determined
from the Balmer decrement and an analysis of the observed SED from the
UV to the optical domains. In other words, it assumes that the young 
star-forming population emitting the UV and Lyman continuum can be wholly 
accounted for in this manner. This method should be correct if our sources do 
not contain highly dust-obscured star-forming regions which are invisible in 
the UV-to-optical range. 
The sky regions where our galaxies are located have been observed in the 
mid-infrared range by {\sl Wide-field Infrared Survey Explorer} ({\sl WISE}). 
Data for all these galaxies but one are 
present in the AllWISE Source 
Catalog\footnote{http://irsa.ipac.caltech.edu/Missions/wise.html.}.
Only the galaxy J0925$+$1403 appears in the list of rejected objects. The 
{\sl WISE} magnitudes of the four studied galaxies are given in 
Table \ref{tab2}. 
The {\sl WISE} colours of our galaxies do not resemble those of
the most nearby star-forming galaxies which have a major contribution from 
stellar emission in the $W1$ and $W2$ bands and are 
characterized by $W1-W2$ $<$ 0.5 mag. Instead, the red $W1-W2$ $>$ 1 of our 
objects imply a considerable contribution of the warm and hot dust to their 
mid-infrared luminosity \citep{I14a}. 

Three out of five galaxies were detected by {\sl WISE} in the $W4$ band at 
22$\mu$m. The estimated ratio of the 22$\mu$m luminosity, which is a 
characteristic of warm dust, to the luminosity absorbed
at 1500\AA\ in these galaxies is of order of unity (Table \ref{tab9}).
As the luminosity of the cold dust emitting in the far-infrared range of 
extreme starbursts (not accessible to {\sl WISE}) is less or comparable
to the luminosity of the warm dust \citep{I14b}, we conclude that the
UV luminosity absorbed in visible star-forming regions of our galaxies is 
consistent with the luminosity emitted by dust in the infrared range.
Optical, near- and mid-infrared observations of other
low-metallicity SFGs with similar properties suggest that 
they are relatively transparent \citep{IT11,IT16}. 
That there is not much dust-obscured hidden
star formation is also implied by the observed thermal free-free cm radio 
emission in dwarf galaxies  
with optical and infrared observations. Its flux density is   
consistent with the value derived from the flux density of the H$\beta$ 
emission line \citep{I14b}, indicating that there is not a considerable amount 
of additional star formation, not seen in the UV range, in our galaxies.

\begin{figure*}
\includegraphics[angle=-90,width=0.47\linewidth]{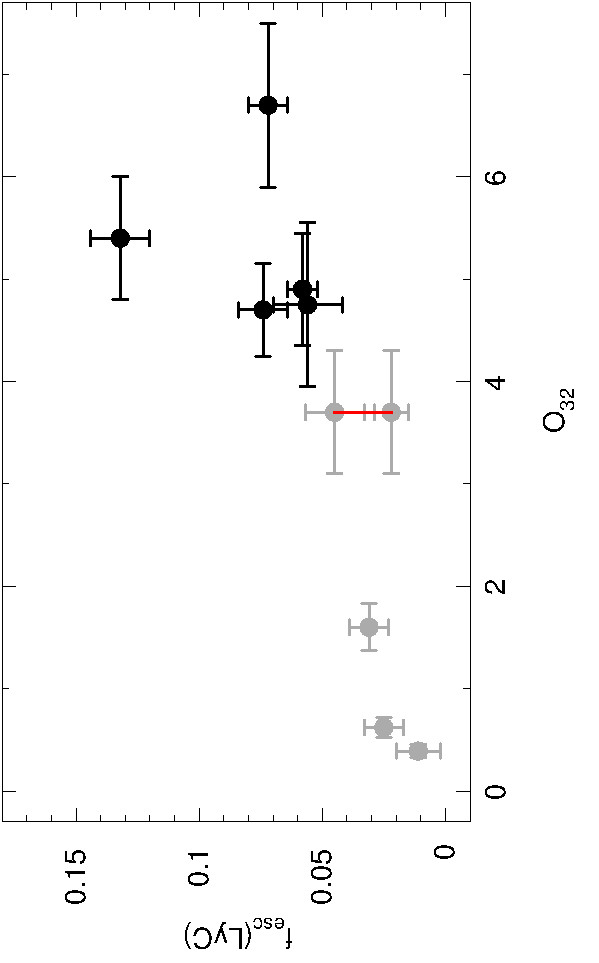}
\caption{Relation between the Lyman continuum escape fraction
$f_{\rm esc}$(LyC) and the [O~{\sc iii}]$\lambda$5007/[O~{\sc ii}]$\lambda$3727 
emission-line ratio in low-redshift LyC leaking galaxies. GP galaxies from
\citet{I16} and this paper are shown by black symbols, while the galaxies from
\citet{L13}, \citet{B14} and \citet{L16} are represented by grey symbols.
Two different measurements of $f_{\rm esc}$(LyC) 
in Tol 1247$-$232 are connected by the red line.
\label{fig14}}
\end{figure*}

\section{Discussion}\label{sec:discuss}

To the best of our knowledge, our experiment is the first where compact 
star-forming galaxies with strong emission lines and high \oiiil/\oiil\ 
ratios are observed in the Lyman continuum.
Furthermore, the entire sample (5 out of 5)
at $z \sim 0.3$ is found to emit Lyman continuum radiation, 
with absolute escape fractions ranging between 6~\% and 13~\%.
Our observations thus significantly expand the sample of Lyman continuum
leakers at low redshift, from four previously known galaxies \citep{L13,B14,L16}
to nine.
The newly discovered sources also show significantly higher LyC escape fractions,
compared to \fesclyc $\approx (1-4.5)$ \% previously found.

Apparently, the selection criteria we adopted are very efficient in finding LyC 
leakers. As such, this finding is probably one of the most important results 
of our study. 

Compactness and a peculiar emission line ratio as selection 
criteria, were previously already suggested to be related to Lyman continuum 
leakage.
\cite{JO13} and \cite{NO14} have proposed that extreme  \oiiil/\oiil\ ratios 
may be due to high escape fractions of ionizing photons and have argued that 
this could be the case for some ``Green Pea" galaxies at $z \sim 0.1-0.3$ or 
in $z \sim 2-3$ LAEs. 
For comparison, \cite{H11} examined so-called Lyman-break analogs (LBAs)
and suggested that a subset of them, those with a compact massive 
($\sim 10^9$ M$_\odot$) dominant central object, could be LyC leakers. 
The LBAs have lower excitation and higher masses compared to our galaxies. 
The leaking galaxy J0921+4509, recently found by \cite{B14} is currently the 
only LBA for which escaping Lyman continuum radiation has been found at the
low level of $\sim$ 1 \%. However, the \oiiil/\oiil\ ratio in this galaxy of 
$\sim$ 0.3 is much lower than that for the GPs in our sample.

In Fig. \ref{fig14} we present the relation between \fesclyc\ and
O$_{32}$ for known low-redshift LyC leaking galaxies.
It shows a trend of increasing \fesclyc\ with increasing O$_{32}$, implying 
that compact low-mass SFGs with high \oiiil/\oiil\ 
ratios may lose a considerable fraction of their LyC emission to the IGM.
The correlation in Figure \ref{fig14} suggests that the high O$_{32}$
criterion plays an important role: it selects out low-metallicity and hence 
low-mass galaxies. 
Our observations \citep[][and this paper]{I16} are the first reported LyC 
observations -- and successful detections -- of ``Green Pea'' galaxies, which 
are a subset of a wider category called luminous compact galaxies (LCGs)
by \cite{I11}. 

At high redshifts, searches for Lyman continuum leakers have been very 
difficult, and numerous candidates have not withstood careful examination 
(see Introduction and references therein). In particular, most candidates
have turned out to be due to chance alignments with foreground objects, as
revealed by {\sl HST} imaging \citep[cf.][]{V10b,V12}.
Currently the most robust high-redshift Lyman continuum leaking galaxy is 
arguably {\em Ion2} at $z=3.2$ \citep{Va15,B16}. Interestingly, this galaxy 
shares many properties with the sources found here (cf.\ Schaerer et al., 
in preparation). In particular, follow-up observations of {\em Ion2} -- 
found from imaging in the Lyman continuum -- have shown a high ratio 
O$_{32}$ $> 10$, again in line with the criterion used to select our targets.

The available data, at high- and low-$z$, and our successful new {\sl HST} 
observations thus all concur to  indicate that compactness and high  O$_{32}$
are promising criteria to identify Lyman continuum leakers.
However, we note that these conditions are necessary but not
sufficient for selecting galaxies with strong Ly$\alpha$ emission
line and escaping LyC emission. For example, the nearby BCD SBS 0335--052E
is characterised by a high O$_{32}$ $\sim$ 10, but its Ly$\alpha$ line is in
absorption \citep{TI97}. The LyC flux in its {\sl FUSE} spectrum
is very low \citep*{T05}, although the data are not of high quality. 
Clearly, larger statistics are needed to verify how good criteria of 
compactness and high O$_{32}$ are for selecting LyC leaking galaxies.
In particular, future observations and further exploration will show 
how the escape fraction varies with \oiiil/\oiil\ ratio, and how it 
depends on the physical parameters of the galaxies.

Although  successful in finding Lyman continuum leaking galaxies, our 
selection criteria as such do not yet explain the {\it physical cause} of the 
observed escape of ionizing photons. Compactness, or more precisely a high 
star formation rate surface density, $\Sigma$, inducing a strong feedback,
rapid outflow etc., may physically be related to leakage of Lyman continuum 
radiation, as suggested
e.g.\ by \citet{H11} and \citet{B14}, and explored further by \citet{Sh16}.
As shown previously in Fig.\  \ref{fig7}, our sources have a very large 
$\Sigma$ both compared to low- and high-redshift sources. Whether they also 
show strong/efficient outflows capable of carving out channels in the ISM or 
blowing it (partly) away, remains to be seen. Further analysis, such as that
of the interstellar absorption lines may shed light on this issue.

\section{Conclusions}\label{summary}

In this paper we present new {\sl Hubble Space Telescope} ({\sl HST}) Cosmic
Origins Spectrograph (COS) observations of four sources in our sample of five 
compact star-forming galaxies (SFGs) with high O$_{32}$ = 
[O~{\sc iii}]$\lambda$5007/[O~{\sc ii}]$\lambda$3727 flux ratios 
$\ga$ 5 at redshifts $z$ $\sim$ 0.3, aiming to study Ly$\alpha$ emission
and escaping Lyman continuum (LyC) radiation.
The first observations of our sample, showing a clear leakage of LyC 
radiation in one galaxy, have been reported in \cite{I16}.
Our main results are as follows:

1. Escaping LyC radiation is detected in all five galaxies. We derive
absolute escape fractions $f_{\rm esc}$(LyC) in the range of $\sim$ 6~\% -- 13~\%,
the highest values found so far in low-redshift SFGs.

2. A double-peaked Ly$\alpha$ emission line, with a small separation between the
blue and red peaks, is detected in the
spectra of all galaxies, as suggested by \cite{V15} for LyC leaking galaxies.
The luminosities and rest-frame equivalent widths of the
Ly$\alpha$ emission line are among the highest found so far for 
Ly$\alpha$ emitters (LAEs) at any redshift. We obtain escape fractions 
$f_{\rm esc}$(Ly$\alpha$) $\sim$ 20~\% -- 40~\%, also among the highest known for 
LAEs.

3. The COS/NUV acquisition images reveal bright star-forming regions in the 
centers of galaxies and an exponential disc at the outskirts with a disc scale 
length in the range $\sim$ 0.6 -- 1.4 kpc, indicating that the galaxies are 
dwarf disc systems. 

4. All five galaxies are characterized by high star-formation rates
SFR $\sim$ 14 -- 50 M$_\odot$ yr$^{-1}$. They have high SFR densities
$\Sigma$ $\sim$ 2 -- 35 $M_\odot$ yr$^{-1}$kpc$^{-2}$, among the highest
known so far for star-forming galaxies at any redshift.
Their stellar masses $M_\star$ are in the range $\sim (0.2-4)\times 10^9$ \msun.
Their metallicities, accurately determined from the optical emission lines 
are in the range $12+\log({\rm O/H})=7.8-8.0$, or $\sim 1/8-1/5$ solar.
These properties are comparable to those of high-redshift star-forming 
galaxies, such as LAEs.

5. From modelling of the UV-to-optical SEDs we constrained the extinction law. 
We find that SEDs reddened with extinction curves by \citet{C89} fit
best the observed COS UV spectra if the selective extinction 
$R(V)$=$A(V)$/$E(B-V)\approx 2.4-2.7$ for the internal galaxy extinction is 
lower than the canonical value of 3.1 and the empirical value of 
4.05 $\pm$ 0.80 for starburst galaxies by \citet{C94}. 
This finding is consistent with the data for the Small Magellanic Cloud with a
similar metallicity.

6. The observations demonstrate that a selection for compact high-excitation
star-forming galaxies showing a high ratio of 
[O~{\sc iii}]$\lambda$5007/[O~{\sc ii}]$\lambda$3727, thus combining various 
criteria suggested in the literature \citep[cf.][]{H11,JO13,NO14}, appears to 
pick up very efficiently sources with escaping Lyman continuum emission. 
Our results should open new and more effective ways to find and explore the 
sources of cosmic reionization in the near future.

\section*{Acknowledgements}

We thank H. Yang for valuable comments.
Based on observations made with the NASA/ESA {\sl Hubble Space Telescope}, 
obtained from the data archive at the Space Telescope Science Institute. 
STScI is operated by the Association of Universities for Research in Astronomy,
Inc. under NASA contract NAS 5-26555. Support for this work was provided by 
NASA through grant number HST-GO-13744.001-A from the Space Telescope Science 
Institute, which is operated by AURA, Inc., under NASA contract NAS 5-26555. 
I.O. acknowledges a grant GACR 14-20666P of the Czech National Foundation. 
Funding for the SDSS and SDSS-II was provided by the Alfred P. Sloan 
Foundation, the Participating Institutions, the National Science Foundation, 
the U.S. Department of Energy, the National Aeronautics and Space 
Administration, the Japanese Monbukagakusho, the Max Planck Society, and the 
Higher Education Funding Council for England. The SDSS was managed by the 
Astrophysical Research Consortium for the Participating Institutions.
Funding for SDSS-III has been provided by the Alfred P. Sloan Foundation, 
the Participating Institutions, the National Science Foundation, and the U.S. 
Department of Energy Office of Science. The SDSS-III web site is 
http://www.sdss3.org/. SDSS-III is managed by the Astrophysical Research 
Consortium for the Participating Institutions of the SDSS-III Collaboration. 
GALEX is a NASA mission  managed  by  the  Jet  Propulsion  Laboratory.
This research has made use of the NASA/IPAC Extragalactic Database (NED) which 
is operated by the Jet  Propulsion  Laboratory,  California  Institute  of  
Technology,  under  contract with the National Aeronautics and Space 
Administration.
This publication makes use of data products from the Wide-field Infrared 
Survey Explorer, which is a joint project of the University of California, 
Los Angeles, and the Jet Propulsion Laboratory/California Institute of 
Technology, funded by the National Aeronautics and Space Administration.






\bsp	
\label{lastpage}
\end{document}